\documentclass[twocolumn,superscriptaddress,prd,aps]{revtex4-2}

\usepackage{graphicx}
\usepackage{dcolumn}
\usepackage{bm}

\usepackage{stmaryrd}   
\usepackage{amsfonts}
\usepackage{mathrsfs}
\usepackage{enumerate}
\usepackage{amsmath}
\usepackage{amsthm}
\usepackage{mathtools}
\usepackage{amssymb}
\usepackage{verbatim}
\usepackage{shuffle}
\usepackage{tikz}
\usepackage{tkz-tab}
\usepackage[cal=boondoxo, calscaled=1.05]{mathalfa}

\newcommand{\ue}{\text{e}}

\begin{document}

\title{Tree- and one-loop-level double copy for the (anti)self-dual sectors of Yang-Mills and gravity}

\author{Daniel Herrera Correa}
\email{daherreraco@unal.edu.co}
\affiliation{Escuela de Matem\'{a}ticas, Universidad Nacional de Colombia Sede Medell\'{i}n, Carrera 65 $\#$ 59A--110, Medell\'{i}n, Colombia}
\author{Cristhiam Lopez-Arcos}
\email{cmlopeza@unal.edu.co}
\affiliation{Escuela de Matem\'{a}ticas, Universidad Nacional de Colombia Sede Medell\'{i}n, Carrera 65 $\#$ 59A--110, Medell\'{i}n, Colombia}
\affiliation{Institute of Physics of the Czech Academy of Sciences \& CEICO Na Slovance 2, 182 21, Prague -- Czech Republic}
\author{Alexander Quintero Vélez}%
\email{aquinte2@unal.edu.co}
\affiliation{Escuela de Matem\'{a}ticas, Universidad Nacional de Colombia Sede Medell\'{i}n, Carrera 65 $\#$ 59A--110, Medell\'{i}n, Colombia}




\date{\today}

\begin{abstract}
By employing the perturbiner method we study the tree- and one-loop-level amplitudes in (anti)self-dual Yang-Mills, focusing on color-kinematics duality and double copy features; they arise naturally even in the fully off-shell case. In particular, we calculate the respective the Kawai-Lewellen-Tye relations for tree-level Berends-Giele currents and color-kinematics master numerators at one loop, both cases for any number of external particles.
\end{abstract}

\maketitle


\section{Introduction}
Feynman diagrams have long been the standard tool for calculations in perturbative quantum field theory with off-shell external particles. However, they are not the most efficient approach for computing scattering amplitudes. This inefficiency stems from the significant combinatorial complexity they impose and the way they obscure the recursive structure of the amplitudes.

The so-called on-shell methods (see \cite{dixon1996, elvang2014, Travaglini_2022, Bern_2007}) introduced a new perspective to amplitude calculations by moving away from the action/diagrammatic approach. This shift not only significantly enhanced computational efficiency but also deepened our understanding of scattering amplitudes. Meanwhile, alternative off-shell methods, though developed with less intensity, have also shown substantial progress due to their richer underlying structure, such as the worldline formalism \cite{Schubert_2001}.

Another perspective introduced by Berends and Giele \cite{Berends:1987me} involves an algebraic framework arising from the analysis of Feynman diagrams. They introduced what is now referred to as a ``current'', a concept that consists of a tree-level diagram with an off-shell leg. This simple idea captures the recursive structure of the amplitudes. More interestingly, one-loop integrands can be obtained by sewing tree-level amplitudes with two off-shell legs \cite{Mahlon_1994}.

It is possible to obtain the Berends-Giele currents directly from the classical equations of motion. The key ingredient is a formal solution called the \emph{perturbiner} expansion \cite{Rosly_1997, rosly1997gravsdperturbiner}, which enables efficient computation of tree-level amplitudes across various theories \cite{Mafra_2015, Lee_2016, Mafra_2016, Mafra_2017, Mizera_2018, Garozzo:2018uzj, Lopez_Arcos_2019, Gomez_2021, Gomez:2021shh, Cho:2021nim, Cho:2022faq}. Its off-shell version serves as a generalization for the Berends-Giele current, utilizing a multiparticle ansatz that resolves the interacting part of the equations without any restrictions on the external particles. Additionally, a sewing procedure can be applied to obtain one-loop off-shell currents, known as one-loop pre-integrands. This method was described in \cite{Gomez:2022dzk} for color-ordered one-loop correlators. More recently, it has been successfully used to calculate one-loop $n$-point correlators in a colorless, non-trivial theory, such as pure gravity \cite{gomez2024}.

There is another paradigm for calculating gravity amplitudes known as the double copy. Originating from string theory as the Kawai-Lewellen-Tye (KLT) duality \cite{Kawai:1985xq}, this concept has been refined into a modern prescription relying on color-kinematics duality for Yang-Mills theory \cite{Bern_2007, Bern_2008, Bern_2010}. The numerators for tree-level gauge amplitudes can be rearranged such that the color and kinematic parts are factorized and follow identical algebraic relations. This allows tree-level graviton amplitudes to be obtained as two copies of the gauge theory kinematic numerators. The extension of this relation to loop level has been conjectured but remains a challenging problem to solve \cite{Bern:2013yya, adamo2022}. To study these dualities, one can examine a simpler sector of Yang-Mills theory, specifically its (anti)self-dual sector, which has been intensely studied over the last few decades \cite{Parkes:1992rz, Bardeen:1995gk, Cangemi:1996pf, Chalmers:1996rq}. This sector provides a particularly fruitful environment for testing various approaches that may help solve this puzzle. Several studies \cite{Monteiro:2011pc, Boels:2013bi} have made significant progress in this regard.

In this work, we use the perturbiner expansion to study the (anti)self-dual sectors of Yang-Mills theory and gravity. At tree-level, we analyze its fully off-shell amplitudes, which exhibit color-kinematics duality even under that condition. At loop level, we obtain one-loop off-shell pre-integrands and then move to the on-shell case to calculate the color-kinematics master numerators. The starting point is the equations of motion of the self-dual sector (see \cite{Campiglia_2021}). The main purpose is to show that the perturbiner expansion and the one-loop pre-integrands provide a complete understanding of the one-loop integrands for these theories, with a powerful recursive prescription that can be applied further to calculate $n$-point one-loop on-shell and off-shell integrands. Moreover, this procedure directly extends the well-known tree-level double-copy structure of amplitudes in self-dual Gravity given by the square of self-dual Yang-Mills, to the one-loop case for these theories.

This paper is organized as follows. In Section \ref{section:sdym-gr}, we introduce the fundamentals for the equations of motion of Yang-Mills and gravity in the (anti)self-dual sectors. Section \ref{section:perturbiners} focuses on the perturbative analysis of (anti)self-dual Yang-Mills theory using perturbiner methods at both tree- and one-loop levels. Section \ref{section:gr-dc} extends this analysis to self-dual gravity, employing its equations of motion and incorporating the double copy relation. In Section \ref{section:concl}, we present our conclusions.

\section{Self-dual and anti-self-dual Yang-Mills and gravity}\label{section:sdym-gr}
Hereafter, we will work with Minkowski spacetime $\mathbb{R}^{1,3}$, employing light-cone coordinates $(u, v, z, \bar{z})$, which are linked to standard Cartesian coordinates $x^{\mu}$ through
\begin{alignat}{2}
\begin{aligned}
    u &= \frac{1}{\sqrt{2}}(x^0+x^3), &\quad v &= \frac{1}{\sqrt{2}}(x^0-x^3), \\
    z &= \frac{1}{\sqrt{2}}(x^1+ix^2), &\quad \bar{z} &= \frac{1}{\sqrt{2}}(x^1-ix^2).
\end{aligned}
\end{alignat}
In these coordinates, the metric has 
the form
\begin{equation}
    \eta = -2 du \otimes dv + 2 dz \otimes d\bar{z}.
\end{equation}
However, to properly deal with the self-dual and anti-self-dual structures, we need to separate the light-cone coordinates into two groups of two-dimensional coordinates. For the self-dual sector, we set $u^{i}=(u,\bar{z})$ and $v^{\alpha} = (v,z)$, whereas for the anti-self-dual sector, we set $u^{i}=(u,z)$ and $v^{\alpha} = (v,\bar{z})$. In either case, the metric for the spacetime may be written in the form
\begin{equation}
    \eta = 2 \eta_{i\alpha} du^{i} \otimes dv^{\alpha}.
\end{equation}
In order to define the (anti)self-dual sectors we also need the tensors from the ``area elements'' arising from these two groupings of coordinates, characterized by
\begin{align}
    \begin{split}
        \Omega^+_{ij} du^{i} \wedge du^{j} &= du \wedge d\bar{z} - d\bar{z} \wedge du, \\
  \Pi^+_{\alpha\beta}dv^{\alpha} \wedge dv^{\beta} &= dv \wedge dz - dz \wedge dv, \\
         \Omega^-_{ij} du^{i} \wedge du^{j} &= du \wedge dz - dz \wedge du, \\
        \Pi^-_{\alpha\beta}dv^{\alpha}\wedge dv^{\beta} &= dv \wedge d\bar{z} - d\bar{z} \wedge dv.
    \end{split}
\end{align}
We will regard $\Omega^{\pm}_{ij}$, and $\Pi^{\pm}_{\alpha\beta}$ as antisymmetric tensors defined in $\mathbb{R}^{1,3}$, where $\Omega^{\pm}_{\alpha\mu}$ and $\Pi^{\pm}_{i\mu}$ are set to zero. Through the use of the inverse of the metric $\eta^{i\alpha}$ for index raising, we can treat these tensors as partial inverses of each other, which means,
\begin{align}
    \begin{split}
        \Omega_{i}^{\pm\alpha} \Pi_{\alpha}^{\pm j} &= \delta_{i}^{j}, \\
        \Pi_{\alpha}^{\pm i}\Omega_{i}^{\pm\beta} &= \delta_{\alpha}^{\beta}.
    \end{split}
\end{align}

Let us now briefly discuss the self-dual and anti-self-dual Yang-Mills equations. The fundamental dynamical variable is a Yang–Mills gauge field $A_{\mu}$ over $\mathbb{R}^{1,3}$, which is valued in the Lie algebra $\mathfrak{g}$ of some compact Lie group. So we can write $A_{\mu}= A_{\mu}^{a}T^{a}$ where the $T^{a}$ are the generators of $\mathfrak{g}$. The corresponding field strengths, $F^{a}_{\mu\nu}$, are defined by 
\begin{align}
  F^{a}_{\mu\nu} = \partial_{\mu} A_{\nu}^{a} - \partial_{\nu} A_{\mu}^{a} - i f^{abc} A_{\mu}^{b} A_{\nu}^{c},  
\end{align}
in which the $f^{abc}$ are the structure constants of $\mathfrak{g}$. The Yang-Mills gauge field $A_{\mu}$ is said to be self-dual (upper sign) or anti-self-dual (lower sign) if its field strengths satisfy
\begin{equation}\label{eq:7}
   i F^{a}_{\mu\nu} = \pm\tfrac{1}{2} \varepsilon_{\mu\nu}^{\phantom{\mu\nu}\rho\sigma} F^{a}_{\rho\sigma},
\end{equation}
where $\varepsilon_{\mu\nu\rho\sigma}$ is the Levi-Civita symbol. For our purposes, it will suffice to limit our treatment using the gauge-fixing condition $A_{u} = A_{\bar{z}}=0$ for the self-dual case and $A_{u}= A_{z} =0$ for the anti-self-dual case. Such gauge-fixing conditions imply that the self-dual and anti-self-dual equations \eqref{eq:7} can be satisfied by setting
\begin{equation}\label{eq:8}
   A^{\pm}_{\alpha} = \Pi_{\alpha}^{\pm i}\partial_{i}\phi^{\pm},
\end{equation}
in which $\phi^{\pm}$ denote $\mathfrak{g}$-valued functions on $\mathbb{R}^{1,3}$, required to satisfy,
\begin{equation}\label{eq:9}
   \Box \phi^{\pm} \pm i \Pi^{\pm ij} [\partial_i\phi^{\pm}, \partial_j\phi^{\pm}] = 0,  
\end{equation}
where $\Box =2\eta^{i \alpha} \partial_{i} \partial_{\alpha}$ is the d'Alembertian operator. We choose equations \eqref{eq:9} as our point of departure for the further constructions.

We now discuss the analogous construction for self-dual and anti-self-dual gravity. In this case, the fundamental dynamical variable is a metric tensor $g_{\mu\nu}$ in some four-dimensional Lorentzian manifold. The corresponding curvature tensor, $R_{\mu\nu\kappa}^{\phantom{\mu\nu\kappa}\lambda}$, is given by
\begin{equation}
R_{\mu\nu\kappa}^{\phantom{\mu\nu\kappa}\lambda} = \partial_{\mu} \Gamma_{\nu\kappa}^{\lambda} - \partial_{\nu} \Gamma_{\mu\kappa}^{\lambda} + \Gamma_{\mu\kappa}^{\rho}\Gamma_{\nu\rho}^{\lambda} - \Gamma_{\nu\kappa}^{\rho}\Gamma_{\mu\rho}^{\lambda},
\end{equation}
where the Christoffel symbols $\Gamma_{\mu\nu}^{\rho}$ are, as usual, defined via the metric as
\begin{equation}
\Gamma_{\mu\nu}^{\rho}= \tfrac{1}{2} g^{\rho\sigma} (\partial_{\mu}g_{\nu\sigma} + \partial_{\nu}g_{\mu\sigma} - \partial_{\sigma}g_{\mu\nu}).  
\end{equation}
The metric tensor $g_{\mu\nu}$ is said to be self-dual (upper sign) or anti-self-dual (lower sign) if its curvature tensor satisfies
\begin{equation}\label{eq:12}
iR_{\mu\nu\kappa}^{\phantom{\mu\nu\kappa}\lambda} = \pm\tfrac{1}{2} \varepsilon_{\mu\nu}^{\phantom{\mu\nu}\rho\sigma}R_{\rho\sigma\kappa}^{\phantom{\rho\sigma\kappa}\lambda}.
\end{equation}
Here we are interested in these conditions for a metric tensor defined in $\mathbb{R}^{1,3}$ of the form $g_{\mu\nu}= \eta_{\mu\nu} + h_{\mu\nu}$, where $h_{\mu\nu}$ is a symmetric tensor field, which can be thought of as the graviton field. And in alignment with our objectives, we will enfore the gauge-fixing condition $h_{\mu u} = h_{\mu \bar{z}} = 0$ for the self-dual case and $h_{\mu u} = h_{\mu z} = 0$ for the anti-self-dual case. This allows the self-dual and anti-self-dual equations \eqref{eq:12} to be satisfied
by setting 
\begin{equation}\label{eq:13}
    h^{\pm}_{\alpha\beta} = \Pi_{\alpha}^{\pm i}\Pi_{\beta}^{\pm j} \partial_{i}\partial_{j} \psi^{\pm},
\end{equation}
where $\psi^{\pm}$ stand for a scalar functions on $\mathbb{R}^{1,3}$, required to satisfy,
\begin{equation}\label{eq:14}
       \Box \psi^{\pm} \mp \tfrac{1}{2}\Pi^{\pm ij}\Pi^{\pm kl}\partial_{i}\partial_{k}\psi^{\pm}\partial_{j}\partial_{l}\psi^{\pm} = 0.
\end{equation}
Once again, equations \eqref{eq:14} serve as the basis for our ongoing constructions.

\section{Multiparticle solutions of self-dual and anti-self-dual Yang-Mills}\label{section:perturbiners}
In this section we turn our attention to the multiparticle solutions to equations \eqref{eq:9}, which, as we have seen, will lead to multiparticle solutions to the self-dual and anti-self-dual equations \eqref{eq:7}. These solutions are both derived via a perturbiner expansion, enabling the recursive generation of fully off-shell versions of Berends-Giele currents. Following this, through the systematic sewing procedure that builds upon and modifies the approach described in \cite{Gomez:2022dzk}, we recursively produce one-loop off-shell currents.

\subsection{Color-stripped perturbiner}
We start by examining the case in which the perturbiner expansions are color-stripped. We keep the notation and conventions of the previous section. 

Consider an infinite set of color indices $(a_p)_{p \geq 1}$ associated with the Lie algebra $\mathfrak{g}$, and an infinite set of massless momentum vectors $(k_p)_{p \geq 1}$ in $\mathbb{R}^{1,3}$.  Let $\mathcal{W}_n$ denote the set of words of length $n$. If $P = p_1p_2\cdots p_n$ is a word in this set, we put $T^{a_P} = T^{a_{p_1}}T^{a_{p_2}} \cdots T^{a_{p_n}}$ and $k_P = k_{p_1} + k_{p_2} + \cdots + k_{p_n}$. Then the color-stripped perturbiner expansions for the equations in \eqref{eq:9} are solution ans\"{a}tze of the form
\begin{equation}
    \phi^{\pm}(x) = \sum_{n \geq 1} \sum_{P \in \mathcal{W}_n} \Phi^{\pm}_P \ue^{ik_P \cdot x} T^{a_P}.
\end{equation}
Upon substituting these expansions back into said equations, and equating coefficients with the same number of generators $T^{a_p}$ on both sides, we find the following recursion relations
\begin{equation}\label{eq:16}
\Phi^{\pm}_{P}=\mp\frac{i}{s_{P}}\sum_{P=QR}\Pi^{\pm ij}k_{iQ}k_{jR}\Phi^{\pm}_{Q}\Phi^{\pm}_{R}.
\end{equation}
Here, $s_p = k_P^2$ is the Mandelstam invariant and the notation $\sum_{P=QR}$ indicates summing over the deconcatenations of the word $P$ into non-empty words $Q$ and $R$.

Now, our primary focus lies in the color-stripped perturbiner expansions for the self-dual and the anti-self-dual Yang-Mills fields $A^{\pm}_{\mu}$. In terms of the above notation, these take the form
\begin{equation}
    A^{\pm}_{\mu}(x) = \sum_{n \geq 1} \sum_{P \in \mathcal{W}_n}  \mathcal{A}^{\pm}_{\mu P} \ue^{ik_P \cdot x} T^{a_P},
\end{equation}
where, in view of the gauge-fixing constraints and \eqref{eq:8}, it is set that $\mathcal{A}^{+}_{uP} = \mathcal{A}^{+}_{\bar{z}P} = 0$, $\mathcal{A}^{-}_{uP} = \mathcal{A}^{-}_{zP} = 0$ and 
\begin{equation}\label{eq:18}
\mathcal{A}^{\pm}_{\alpha P} = i \Pi_{\alpha}^{\pm i} k_{iP} \Phi^{\pm}_P.
\end{equation}
If \eqref{eq:16} is applied to \eqref{eq:18}, after a little algebra it is easy to see that the latter may be arranged as
\begin{align}\label{eq:19}
\begin{split}
    \mathcal{A}^{\pm}_{\alpha P} &= \pm\frac{1}{s_P} \sum_{P=QR} \{\eta^{i\beta}\mathcal{A}^{\pm}_{\beta Q}  k_{iR}  \mathcal{A}^{\pm}_{\alpha R}-\eta^{i\beta}\mathcal{A}^{\pm}_{\beta R}  k_{iQ}  \mathcal{A}^{\pm}_{\alpha Q}\} \\
   &= \pm\frac{1}{s_P} \sum_{P=QR} \{(\mathcal{A}^{\pm}_{Q} \cdot k_R)  \mathcal{A}^{\pm}_{\alpha R}-(\mathcal{A}^{\pm}_{R} \cdot k_Q)  \mathcal{A}^{\pm}_{\alpha Q}\}.
\end{split}
\end{align}
From this, in particular, follows the shuffle constraint $\mathcal{A}^{\pm}_{\alpha P \shuffle Q} =0$, with $P \shuffle Q$ yielding the sum over all possible shuffles between the words $P$ and $Q$.

It is useful at this stage to review, briefly, one of the key points in \cite{Gomez:2022dzk}. If the on-shell condition for the single particle states is dropped, the equations in \eqref{eq:16} define  off-shell recursions that only solve the equations in \eqref{eq:9} at the multiparticle level. As a result, the recursion relations in \eqref{eq:19} solve the interacting part of the self-dual and anti-self-dual equations while keeping the single-particle states off-shell. 
Put another way, the quantities $ \mathcal{A}^{\pm}_{\mu P}$ can be interpreted as off-shell color-stripped Berends-Giele currents for the self-dual and anti-self-dual Yang-Mills fields $A^{\pm}_{\mu}$.

To round off our discussion, it is essential to examine how the off-shell color-stripped Berends-Giele currents are linked to the scattering amplitudes in the self-dual and anti-self-dual sectors of Yang-Mills theory. Referring once more to insights from \cite{Gomez:2022dzk}, this connection is established through a direct extension of the Berends-Giele methodology. However, before we proceed, it is imperative to highlight a key point:~the self-dual and anti-self-dual theories described by the equations in \eqref{eq:9} correspond, respectively, to the positive and negative helicity sectors of Yang-Mills theory; see, for example, \cite{Parkes:1992rz,Bardeen:1995gk,Chalmers:1996rq,Cangemi:1996pf}. With this point in mind, and writing $\varepsilon^{\pm}_p = \mathcal{A}^{\pm}_{p}$ for the single index color-stripped Berends-Giele currents, the off-shell tree-level double-partial amplitudes are determined using the formula
\begin{equation}\label{eq:20}
  \begin{split}
       A^{\mathrm{tree}}(1^{h_{1}}P^{h_{1}}n^{h_n}) = s_{1P} \mathcal{A}^{h_{1}}_{1P} \cdot \varepsilon^{h_n}_{n},
  \end{split}  
\end{equation}
where the word $P$ is a permutation of $23 \cdots (n-1)$, $-k_n = k_1 + k_P$ is fixed by momentum conservation and the superscripts $h_{1}$ and $h_n$ indicate the helicity.

To demonstrate the application of \eqref{eq:20}, we briefly consider two examples. First, we consider an $n$ gluon configuration where all the particles have the same helicity. Thus \eqref{eq:20} can be expressed as
\begin{equation}
A^{\mathrm{tree}}(1^{\pm}P^{\pm}n^{\pm}) =  s_{1P} \eta^{i \alpha} \mathcal{A}^{\pm}_{\alpha 1P} \varepsilon^{\pm}_{i n}.
\end{equation}
But the gauge-fixing constraints tell us that, regardless of the grouping of the light-cone coordinates, $ \varepsilon^{\pm}_{i n} = 0$. In this way we find that
\begin{equation}
A^{\mathrm{tree}}(1^{\pm}P^{\pm}n^{\pm}) = 0.
\end{equation}
This is a well-known result that has been reported in various works \cite{Rosly:1996vr,Cangemi:1996rx,Monteiro:2011pc,Boels:2013bi,Chattopadhyay:2022iwk}. Let us now consider an $n$ gluon configuration in which all but one particle have the same helicity. This configuration, in fact, is going to throw the only non-zero tree-level amplitudes in the theory, the so-called MHV amplitudes with two gluons of $\pm$ helicity and one gluon of $\mp$ helicity. Use of the prescription \eqref{eq:20} yields
\begin{equation}
A^{\mathrm{tree}}(1^{\pm}P^{\pm}n^{\mp}) = s_{1P} \eta^{i \alpha} \mathcal{A}^{\pm}_{\alpha 1P} \varepsilon^{\mp}_{i n},
\end{equation}
which, by the grouping of the light-cone coordinates, may be written more compactly as
\begin{equation}\label{eq:24}
A^{\mathrm{tree}}(1^{\pm}P^{\pm}n^{\mp}) =  s_{1P}  \mathcal{A}^{\pm}_{1P} \cdot \varepsilon^{\mp}_{n}.
\end{equation}
The claim is that these tree-level amplitudes all vanish if $n > 3$. To see how this comes about, we resort to the spinor-helicity formalism in which the single index color-stripped Berends-Giele currents are expressed as
\begin{equation}\label{eq:25}
    \varepsilon_{\mu p}^{+} = - \frac{\langle r \vert \gamma_{\mu} \vert k_p ]}{\langle r k_p \rangle } , \quad  \varepsilon_{\mu p}^{-} = - \frac{\langle k_p \vert \gamma_{\mu} \vert r ]}{[ r k_p ] } ,
\end{equation}
where $r$ is an arbitrary massless reference momentum satisfying $k_p \cdot r \neq 0$. Assume, then, that $n > 3$ and let us write $P=p_2 p_3 \cdots p_{n-1}$. Unpacking what equation \eqref{eq:24} conveys, taking into account the recursion relations \eqref{eq:19}, one sees that the tree-level amplitude $A^{\mathrm{tree}}(1^{\pm}P^{\pm}n^{\mp})$ will have numerators that are combinations of monomials in $\varepsilon_{p_i}^{\pm} \cdot \varepsilon_n^{\mp}$ for $2 \leq i \leq n-1$. When expressing each $\varepsilon_{p_i}^{\pm} \cdot \varepsilon_n^{\mp}$ in spinor-helicity notation, it becomes proportional to $\langle k_{p_i} r \rangle [k_n r]$. We now crucially exploit our hypothesis that $n > 3$ by choosing the reference momentum $r$ to be equal to $k_n$, in which case the latter expression vanishes. Such a choice is inadmissible when $n = 3$. This key observation allows us to conclude that $A^{\mathrm{tree}}(1^{\pm}P^{\pm}n^{\mp}) = 0$, as we wished to show. We are thus left with only the amplitudes $A^{\mathrm{tree}}(1^{\pm}2^{\pm}3^{\mp})$. Appealing again to \eqref{eq:24} and \eqref{eq:19}, it is easily seen that
\begin{equation}
    A^{\mathrm{tree}}(1^{\pm}2^{\pm}3^{\mp})=\varepsilon_1^{\pm} \cdot k_2 \varepsilon_2^{\pm} \cdot \varepsilon_3^{\mp} - \varepsilon_2^{\pm} \cdot k_1\varepsilon_1^{\pm} \cdot \varepsilon_3^{\mp}.
\end{equation}
By means of the expressions given in \eqref{eq:25} we then infer that
\begin{equation}\label{eq:27}
    A^{\mathrm{tree}}(1^{+}2^{+}3^{-})=\frac{[12]^3}{[23][31]},
\end{equation}
and, similarly,
\begin{equation}\label{eq:28}
    A^{\mathrm{tree}}(1^{-}2^{-}3^{+})=\frac{\langle 12 \rangle^3}{\langle 23 \rangle \langle 31 \rangle}.
\end{equation}
These expressions match those found in existing literature; see, for instance,  \cite{Rosly:1996vr,Cangemi:1996rx,Monteiro:2011pc,Boels:2013bi,Chattopadhyay:2022iwk}.

\subsection{Color-dressed perturbiner}
We shall now turn to the scenario where the perturbiner expansions are color-dressed. Their treatment can be conducted in a manner akin to that of the color-stripped perturbiner expansions.

The color-dressed perturbiner ans\"{a}tze for the solutions of the equations in \eqref{eq:9} are as follows:
\begin{equation}
    \phi^{\pm a}(x) = \sum_{n \geq 1} \sum_{P \in \mathcal{OW}_n} \Phi^{\pm a}_P \ue^{ik_P \cdot x}.
\end{equation}
Here $\mathcal{OW}_n$ denotes the set of words $P=p_1p_2 \cdots p_n$ of length $n$ with $p_1 < p_2 < \cdots < p_n$. Plugging these ans\"{a}tze back into said equations, one arrives at the  recursion relations
\begin{equation}\label{eq:30}
\Phi^{\pm a}_{P}=\mp\frac{i}{s_{P}}\sum_{P=Q \cup R}f^{abc}\Pi^{\pm ij}k_{iQ}k_{jR}\Phi^{\pm b}_{Q}\Phi^{\pm c}_{R}.
\end{equation}
In this expression the notation $P=Q \cup R$ instructs to distribute the letters of the ordered words $P$ into non-empty ordered
words $Q$ and $R$.

Once again, our consideration centers on the color-dressed perturbiner expansions for the self-dual and anti-self-dual Yang-Mills fields $A^{\pm}_{\mu}$. Building upon the previously introduced notation, we express these expansions as
\begin{equation}
    A^{\pm a}_{\mu}(x) = \sum_{n \geq 1} \sum_{P \in \mathcal{OW}_n}  \mathcal{A}^{\pm a}_{\mu P} \ue^{ik_P \cdot x},
\end{equation}
where, in accordance with the gauge-fixing constraints and the relation given in equation \eqref{eq:8}, it is required that $\mathcal{A}^{+a}_{u P} = \mathcal{A}^{+a}_{\bar{z} P} = 0$, $\mathcal{A}^{-a}_{u P} = \mathcal{A}^{-a}_{z P} = 0$ and 
\begin{equation}\label{eq:32}
\mathcal{A}^{\pm a}_{\alpha P} = i \Pi_{\alpha}^{\pm i} k_{iP} \Phi^{\pm a}_P.
\end{equation}
By substituting the relation from equation \eqref{eq:30} into equation \eqref{eq:32}, after some algebraic manipulation, it becomes apparent that the latter can be recast as 
\begin{equation}\label{eq:33}
    \mathcal{A}^{\pm a}_{\alpha P} =\pm\frac{1}{s_P} \sum_{P=Q \cup R} f^{abc}\{(\mathcal{A}^{\pm b}_{Q} \cdot k_R)  \mathcal{A}^{\pm c}_{\alpha R}-(\mathcal{A}^{\pm c}_{R} \cdot k_Q)  \mathcal{A}^{\pm b}_{\alpha Q}\}.
\end{equation}
As mentioned earlier, disregarding the on-shell condition for the single-particle states allows the quantities $ \mathcal{A}^{\pm a}_{\mu P}$ to be interpreted as off-shell color-dressed Berends-Giele currents for the self-dual and anti-self-dual Yang-Mills fields $A^{\pm}_{\mu}$.

At this juncture, it is instructive to understand the relationship between the off-shell color-stripped and color-dressed Berends-Giele currents. To this end, for each bracketed word $l[P]=l[p_1 p_2 \cdots p_n]$ of lenght $n$, we let $c^{a}_{l[P]}$ denote the product of color factors determined by 
\begin{equation}
    c^{a}_{l[P]}= f^{a_{p_1} a_{p_2} b} f^{b a_{p_3} c} \cdots f^{d a_{p_{n-1}} e} f^{e a_{p_{n}} a},
\end{equation}
with the understanding that $c^{a}_{p}=\delta^{a}_{\phantom{a}a_p}$. We also set 
\begin{equation}
    c^{a}_{[l[P],l[Q]]}=f^{abc}  c^{b}_{l[P]} c^{c}_{l[Q]}
\end{equation}
for every pair of bracketed words $l[P]$ and $l[Q]$. By following the same arguments as those presented in \S13 of \cite{Escudero:2022zdz}, it is then possible to show that
\begin{equation}\label{eq:36}
    \mathcal{A}^{\pm a}_{\alpha 12 \cdots (n-1)} = \sum_{P} c^{a}_{l[P]}  \mathcal{A}^{\pm}_{\alpha 1P},
\end{equation}
where the sum over $P$ is taken over all permutations of $23 \cdots (n-1)$.

It seems now natural to consider the amplitudes associated with the self-dual and anti-self-dual sectors of Yang-Mills theory, which are tied to the off-shell color-dressed Berends-Giele currents. The formulas for calculating such amplitudes reveal themselves as straightforward extensions of the ones explored in the context of the color-stripped case. Indeed, if we decompose the color and kinematic degrees of freedom of the single index color-dressed Berends-Giele currents as $\mathcal{A}^{\pm a}_{\alpha p} = \delta^{a a_{p}} \varepsilon^{\pm}_{\alpha p}$, the full amplitude for the scattering of $n$ gluons can be determined by using
\begin{align}
    \begin{split}
        &\mathcal{A}^{\mathrm{tree}}_n(1^{h_1},2^{h_1},\dots,(n-1)^{h_1},n^{h_n}) \\
        &\qquad \qquad \qquad = s_{12 \cdots (n-1)} \mathcal{A}^{h_1 a_n}_{12 \cdots (n-1)} \cdot \varepsilon^{h_n}_{n},
    \end{split}
\end{align}
where again we assume momentum conservation. By referring to \eqref{eq:36} and the definition \eqref{eq:20} it is easy to see that the latter is expressible in the form
\begin{align}
   \begin{split}
       &\mathcal{A}^{\mathrm{tree}}_n(1^{h_1},2^{h_1},\dots,(n-1)^{h_1},n^{h_n}) \\
       &\qquad \qquad \qquad = \sum_{P} c_{l[P]}^{a_n} A^{\mathrm{tree}}(1^{h_1}P^{h_1}n^{h_n}).
   \end{split} 
\end{align}
From this formula, combined with the analysis in the preceding section, we arrive at the conclusion that all amplitudes vanish identically, except for $\mathcal{A}^{\mathrm{tree}}_3(1^{+},2^{+},3^{-})$ and $\mathcal{A}^{\mathrm{tree}}_3(1^{-},2^{-},3^{+})$, which, according to \eqref{eq:27} and \eqref{eq:28}, take the forms
\begin{equation}
    \mathcal{A}^{\mathrm{tree}}_3(1^{+},2^{+},3^{-})=f^{a_1a_2a_3}\frac{[12]^3}{[23][31]},
\end{equation}
and
\begin{equation}
    \mathcal{A}^{\mathrm{tree}}_3(1^{-},2^{-},3^{+})=f^{a_1a_2a_3}\frac{\langle 12 \rangle^3}{\langle 23 \rangle \langle 31 \rangle}.
\end{equation}
Just as before, these results are consistent with those reported in various studies, such as \cite{Monteiro:2011pc,Boels:2013bi,Chattopadhyay:2022iwk}.

\subsection{Color-kinematics duality}\label{sec:IIIC}
We will now explore how color-kinematics duality for the self-dual and anti-self-dual sectors of Yang-Mills theory. This is made manifest for the off-shell color-dressed Berends-Giele currents obtained in the preceding section. In essence, we will follow the treatment of \S7 of \cite{Escudero:2022zdz}.

Our starting point involves the decomposition of the color and kinematic degrees of freedom of the single index color-dressed Berends-Giele current $\mathcal{A}^{\pm a}_{\alpha p} = \delta^{a a_{p}} \varepsilon^{\pm}_{\alpha p}$. Building on this, let us consider the infinite-dimensional Lie algebra $\mathfrak{g}'^{\pm}$ that is spanned by the $\varepsilon^{\pm}_{\alpha p}$ and whose Lie bracket is defined by
\begin{equation}\label{eq:41}
    [\varepsilon^{\pm}_p,\varepsilon^{\pm}_q]_{\alpha} = (\varepsilon^{\pm}_p \cdot k_q) \varepsilon^{\pm}_{\alpha q} - (\varepsilon^{\pm}_q \cdot k_p) \varepsilon^{\pm}_{\alpha p}.
\end{equation}
Subsequently, for any bracketed word $l[P]=l[p_1 p_2 \cdots p_n]$ of lenght $n$, we define the quantities $\varepsilon^{\pm}_{\alpha l[P]}$ by
\begin{equation}\label{eq:42}
    \varepsilon^{\pm}_{\alpha l[P]} = [[\dots,[[\varepsilon^{\pm}_{p_1},\varepsilon^{\pm}_{p_2}],\varepsilon^{\pm}_{p_3}],\dots],\varepsilon^{\pm}_{p_n}]_{\alpha}.
\end{equation}
We further put
\begin{equation}
    \varepsilon^{\pm}_{\alpha [l[P],l[Q]]}=[ \varepsilon^{\pm}_{l[P]}, \varepsilon^{\pm}_{l[Q]}]_{\alpha}
\end{equation}
for every pair of bracketed words $l[P]$ and $l[Q]$. Next, we need to invoke a certain combinatorial gadget, termed the ``color-dressed Berends-Giele map'' introduced in \cite{Ahmadiniaz:2021ayd}, that allows us to keep track of the bracketed words obtained by iterated recursions of \eqref{eq:33}.  It is defined as the map $b_{\mathrm{cd}}$ acting on ordered words and determined recursively by
\begin{align}\label{eq:44}
    \begin{split}
        b_{\mathrm{cd}}(p) &= p, \\
        b_{\mathrm{cd}}(P) &= \frac{1}{2s_P} \sum_{P = Q \cup R} [ b_{\mathrm{cd}}(Q), b_{\mathrm{cd}}(R)].
    \end{split}
\end{align}
Finally, as a matter of notation,  given two arbitrary labelled objects $U_P$ and $V_P$, we define the replacement of ordered words by the
product of such objects as
\begin{equation}
    \llbracket U \otimes V \rrbracket \circ P = U_P V_P.
\end{equation}
With this understanding, it can be shown that the recursion relations in \eqref{eq:33} can be expressed in the form
\begin{equation} \label{eq:46}
    \mathcal{A}^{\pm a}_{\alpha P} = (\pm 1)^{\lvert P \rvert -1} \llbracket c^{a} \otimes \varepsilon^{\pm}_{\alpha} \rrbracket \circ  b_{\mathrm{cd}}(P).
\end{equation}
Keeping in view \eqref{eq:44}, this amounts to saying that the generalised Jacobi identities associated to the color factors $c^{a}_{l[P]}$ are also satisfied by the quantities $\varepsilon^{\pm}_{\alpha l[P]}$, which can be regarded as kinematic numerators. Given that these numerators are constructed using the structure constants of the infinite-dimensional Lie algebra $\mathfrak{g}'^{\pm}$, we can conclude that $\mathfrak{g}'^{\pm}$ represents a specific realization of the ``kinematic Lie algebra'' underlying the duality between color and kinematics in both the self-dual and anti-self-dual sectors of Yang-Mills theory. 

\subsection{One-loop integrands}
We now endeavor to determine one-loop integrands in the self-dual and anti-self-dual sectors of Yang-Mills theory from the respective off-shell perturbiner expansions. This task is undertaken by applying an algorithm that is a variant of the one used in the sewing procedure outlined in \cite{Gomez:2022dzk}.

We start with the off-shell color-stripped Berends-Giele currents $\mathcal{A}^{\pm}_{\alpha l P}$, in which the single-particle label $l$ plays a distinguished role. Through the application of equation \eqref{eq:19}, we can explicitly express these currents by factoring out the single index color-stripped Berends-Giele current $\mathcal{A}^{\pm}_{l}=\varepsilon^{\pm}_l$ as
\begin{equation}
    \mathcal{A}^{\pm}_{\alpha l P} =  \eta^{i\beta}\mathcal{J}^{\pm}_{i \alpha  P} \varepsilon^{\pm}_{\beta l} ,
\end{equation}
in which
\begin{align}\label{eq:48}
   \begin{split}
    \mathcal{J}^{\pm}_{i \alpha  P} &=\pm\frac{1}{s_{l P}}\bigg\{ \eta_{i\beta}\mathcal{A}^{\pm}_{\gamma P}(\delta^{\gamma}_{\alpha}k^{\beta}_{P}-\delta^{\beta}_{\alpha}\eta^{j\gamma}k_{jl} )  \\
    &\quad + \sum_{P=QR}\mathcal{J}^{\pm}_{i\beta Q}\mathcal{A}^{\pm}_{\gamma R}(\delta^{\gamma}_{\alpha}\eta^{j\beta}k_{j R}-\delta^{\beta}_{\alpha}\eta^{j\gamma}k_{jlQ}) \bigg\}.        
\end{split} 
\end{align}
The currents $\mathcal{J}^{\pm}_{i \alpha  P}$ are what in the terminology of \cite{Gomez:2022dzk} one would call one-loop pre-integrands for self-dual or anti-self-dual gluon fields. 

To proceed, we need to establish some terminology and notation  to facilitate the combinatorial description of the one-loop self-dual and anti-self-dual integrands. Given a word $P$ and a deconcatenation $P=QR$, we will borrow the terminology of \cite{Gomez:2022dzk} and define the cyclic completion of the permutation as $[QR]$. Moreover, since we would need to avoid the case when the length of $R$ is one for the on-shell one-loop integrand, we will further restrict the cyclic completion and denote a deconcatenation $(QR)$ for the elements of the cyclic completion $[QR]$ such that the length of $Q$ is equal or greater that of $R$, and the length of $R$ is at least two. We will refer to $(QP)$ as the restricted cyclic completion.

To express the one-loop integrands, we, in the first place, take the loop momentum to be $\ell = k_l$. Then, in view of \eqref{eq:48}, the one-loop pre-integrands $\mathcal{J}^{\pm}_{i \alpha P}$ may be regarded as functions of $\ell$. In more detail, 
\begin{eqnarray}\label{eq:49}
\mathcal{J}^{\pm}_{i \alpha P}(\ell) &=& \pm\frac{1}{(\ell + k_P)^2}\bigg\{ \eta_{i\beta}\mathcal{A}^{\pm}_{\gamma P}(\delta^{\gamma}_{\alpha}k^{\beta}_{P}-\delta^{\beta}_{\alpha}\eta^{j\gamma}\ell_j )\notag\\ &&+ \sum_{P=QR}\mathcal{J}^{\pm}_{i\beta Q}(\ell)\mathcal{A}^{\pm}_{\gamma R}(\delta^{\gamma}_{\alpha}\eta^{j\beta}k_{j R}\notag\\
&&\hspace{1.3cm}-\delta^{\beta}_{\alpha}\eta^{j\gamma}(\ell + k_Q)_j)\bigg\}.
\end{eqnarray}
The first term in \ref{eq:49} represents a tadpole after sewing. In order to define the integrand, we need to introduce an useful notation distinguishing a preintegrand with and without tadpole. Therefore, we may define a preintegrand without tadpole as
\begin{eqnarray}
  \widetilde{\mathcal{J}}^{\pm}_{i \alpha P}(\ell) &=& \pm\frac{1}{(\ell + k_P)^2}\bigg\{ \sum_{P=QR}\mathcal{J}^{\pm}_{i\beta Q}(\ell)\mathcal{A}^{\pm}_{\gamma R}(\delta^{\gamma}_{\alpha}\eta^{j\beta}k_{j R}\notag\\
  &&\hspace{2.2cm}-\delta^{\beta}_{\alpha}\eta^{j\gamma}(\ell + k_Q)_j)\bigg\}.
\end{eqnarray}    
while the tadpole itself is written as 
\begin{equation}
    \hat{\mathcal{J}}^{\pm}_{i \alpha P}(\ell) =\pm\frac{1}{(\ell + k_P)^2} \eta_{i\beta}\mathcal{A}^{\pm}_{\gamma P}(\delta^{\gamma}_{\alpha}k^{\beta}_{P}-\delta^{\beta}_{\alpha}\eta^{j\gamma}\ell_j )
\end{equation}
Within this terminology, we can define the off-shell one-loop integrand for a word $P$ by taking the cyclic completion of the deconcatenations of $P$ in all passible words $QR$, denoted by $[QR]$, as explained in \cite{Gomez:2022dzk}, as shown below
\begin{equation}
\label{eq:52}
    \bar{I}^{\text{$1$-loop}}_{P}(\ell) =\sum_{P=[QR]} \eta^{i\beta}\eta^{j\alpha}\mathcal{J}_{i\alpha Q}^{\pm}(\ell)\mathcal{J}_{j\beta R}^{\text{tp}\pm}(\ell+k_{Q})
\end{equation}
We are interested in understanding on-shell one-loop integrands.  In order to obtain the on-shell one-loop integrand, it is necessary that we eliminate the tadpole and external leg bubbles coming from some of the contributions in (\ref{eq:52}). To achieve it, we may use the cyclic deconcatenations of restricted length $(QR)$ to define one-loop on-shell integrand for a word $Pn$ by
\begin{eqnarray}
\label{eq:53}
    I^{\text{$1$-loop}}_{Pn}(\ell) &=&\eta^{i\beta}\eta^{j\alpha}\widetilde{\mathcal{J}}_{i\alpha P}^{\pm}(\ell)\hat{\mathcal{J}}_{j\beta n}^{\pm}(\ell+k_{P})\\
    &&+ \sum_{P=(Q\,Rn) }\eta^{i\beta}\eta^{j\alpha}\mathcal{J}_{i\alpha Q}^{\pm}(\ell)\hat{\mathcal{J}}_{j\beta Rn}^{\pm}(\ell+k_{Q})\notag
\end{eqnarray}
This formula doesn't suffer from tadpoles or redundant contributions. 

Here, the momentum conservation $k_P = 0$ and the on-shell transversality condition $\varepsilon^{\pm}_p \cdot k_p = 0$ for every single index Berends-Giele current are understood. Analyzing the structure of the  contraction if two preintegrand as appear in Eq. (52) is of great use to understand how the explicit contributions appear. Therefore,
\begin{widetext}
\begin{align}
\label{eq:54}
    \begin{split}
        &\eta^{i\beta}\eta^{j\alpha}\mathcal{J}^{\pm}_{i\alpha Q}(\ell)\hat{\mathcal{J}}^{\pm}_{j\beta R} (\ell +k_Q)= \frac{\eta^{i\beta}\eta^{j\alpha}}{\ell^2 (\ell + k_Q)^2} \bigg\{  \eta_{i \beta'} \mathcal{A}^{\pm}_{\gamma' Q}(\delta^{\gamma'}_{\alpha} k^{\beta'}_{Q} - \delta^{\beta'}_{\alpha} \eta^{k \gamma'} \ell_k)\\ 
        &+ \sum_{Q=ST} \mathcal{J}^{\pm}_{i\beta' S}(\ell) \mathcal{A}^{\pm}_{\gamma' T} (\delta^{\gamma'}_{\alpha} \eta^{k\beta'} k_{kT} - \delta^{\beta'}_{\alpha}\eta^{k\gamma'} (\ell + k_S)_k ) \bigg\}
         \times \bigg\{ \eta_{j \beta''} \mathcal{A}^{\pm}_{\gamma'' R}(\delta^{\gamma''}_{\beta} k^{\beta''}_{R} - \delta^{\beta''}_{\beta} \eta^{k \gamma''} (\ell+k_Q)_k)  \bigg\}.
    \end{split}
\end{align}
\end{widetext}

To simplify this expression further, notice that when together the first terms of each factor give, after some rewriting, a term of the form
\begin{equation}
\begin{aligned}\label{eq:50}
        &\frac{1}{\ell^2 (\ell + k_Q)^2}\{\mathcal{A}^{\pm}_Q \cdot k_R \ \mathcal{A}^{\pm}_R \cdot k_R  - \mathcal{A}^{\pm}_Q \cdot \ell \ \mathcal{A}^{\pm}_R \cdot k_R \\
        &\:- \mathcal{A}^{\pm}_Q \cdot k_Q \ \mathcal{A}^{\pm}_R \cdot (\ell+k_Q) + 4\mathcal{A}^{\pm}_Q \cdot \ell \ \mathcal{A}^{\pm}_R \cdot (\ell+k_Q) \}.
\end{aligned}
\end{equation}
But the transversality condition for the single index Berends-Giele currents $\varepsilon^{\pm}_p$ translates into a transversality condition for the color-stripped Berends-Giele currents $\mathcal{A}^{\pm}_Q$ and $\mathcal{A}^{\pm}_R$, and moreover, momentum conservation tells us that $k_Q = - k_R$. In light of this, the only nonzero contribution in \eqref{eq:50} is
\begin{equation}
    \frac{4\mathcal{A}^{\pm}_Q \cdot \ell \ \mathcal{A}^{\pm}_R \cdot \ell}{\ell^2 (\ell + k_Q)^2} .
\end{equation}
By following the same reasoning, one can establish that the contributions  in \eqref{eq:50} coming from terms contracted with the factors $\delta^{\gamma'}_{\alpha} \eta^{k\beta'} k_{kT}$, $\delta^{\gamma''}_{\beta} k^{\alpha}_{R}$, $\delta^{\gamma'}_{\alpha} k^{\beta}_{Q}$, and $\delta^{\gamma''}_{\beta} \eta^{k\beta''} k_{kV}$ will all vanish. Hence, the contributions of the one-loop integrand may be written in full as
\begin{widetext}
\begin{align}\label{eq:57}
    \begin{split}
        \eta^{i\beta}\eta^{j\alpha}\mathcal{J}^{\pm}_{i\alpha Q}(\ell)\hat{\mathcal{J}}^{\pm}_{j\beta R} (\ell +k_Q)=  \frac{1}{\ell^2 (\ell + k_Q)^2} \bigg\{& 4\mathcal{A}^{\pm}_Q \cdot \ell \ \mathcal{A}^{\pm}_R \cdot \ell + \sum_{Q=ST}\eta^{i\beta} \mathcal{J}^{\pm}_{i\beta S}(\ell) 
\mathcal{A}^{\pm}_{T} \cdot (\ell + k_S) \ \mathcal{A}^{\pm}_{R} \cdot  (\ell+k_Q) \bigg\}.
    \end{split}
\end{align}
\end{widetext}
This already give us  and idea of the recursive structure of the one-loop integrand for the on-shell situation. 

Let us offer some remarks to clarify the meaning of this formula and its accuracy in generating the expected Feynman diagrams. For this, let us focus again on each summand of \eqref{eq:53}, that is, on each of the contraction $\eta^{i\beta}\eta^{j\alpha}\mathcal{J}^{\pm}_{i\alpha Q}(\ell)\hat{\mathcal{J}}^{\pm}_{j\beta R} (\ell +k_Q)$. It is not too difficult to check that applying the recursion implied in \eqref{eq:49} to \eqref{eq:57} a total of $k-1$ times yields a deconcatenation of $P = (QR)$ into $k$ subwords $P_1,\dots,P_k$ whose associated terms will involve contractions of Berends-Giele currents $\mathcal{A}^{\pm}_{P_s}$ with the loop momentum $\ell$, possibly shifted by $k_{P_i} + \cdots + k_{P_j}$ for $i < j$. We will designate this contribution to the one-loop integrand as $I^{\text{$k$-gon}}_{\ell}(P^{\pm}_1 \vert \cdots \vert P^{\pm}_k)$, since it corresponds exactly to a $k$-gon diagram. For instance, when $k=2$, it follows at once from what has been said that
\begin{equation}
\label{eq:58}
I^{\text{$2$-gon}}_{\ell}(Q^{\pm}\vert R^{\pm}) = \frac{4\mathcal{A}^{\pm}_Q \cdot \ell \ \mathcal{A}^{\pm}_R \cdot \ell}{\ell^2 (\ell + k_Q)^2}.
\end{equation}
Given these considerations, if $n$ denotes the length of $P$, the formulas for the one-loop self-dual and anti-self-dual integrands become
\begin{equation}
I^{\text{$1$-loop}}_P(\ell) = \sum_{k=2}^{n} \sum_{P=P_1 \cdots P_k} I^{\text{$k$-gon}}_{\ell}(P^{\pm}_1 \vert \cdots \vert P^{\pm}_k).
\end{equation}
At this point, it is worth highlighting that each subword $P_i$ contributing to $I^{\text{$k$-gon}}_{\ell}(P^{\pm}_1 \vert \cdots \vert P^{\pm}_k)$ can itself be deconcatenated as $P_i = P_i^1 P_i^2 \cdots P_i^{m_i}$. This deconcatenation leads us to associate a tree with each $P_i$, which branches into its subsubwords $P_i^j$. In this way, the contribution $I^{\text{$k$-gon}}_{\ell}(P^{\pm}_1 \vert \cdots \vert P^{\pm}_k)$ can be visually represented by a $k$-gon diagram.

To illustrate the whole construction, we pursue the analysis for the four point one-loop integrands $I^{\text{$1$-loop}}_4(\ell)$ by applying directly the prescription in Eq. (\ref{eq:53}). In this case, we encounter only two possible deconcatenations coming from the cyclic completion, which are $(12,34)$ and $(23,41)$. This led us to consider 
\begin{equation}
\label{eq:62}
    \begin{split}
        I^{\text{$1$-loop}}_{1234}(\ell) &=\eta^{i\beta}\eta^{j\alpha}\widetilde{\mathcal{J}}_{i\alpha 123}^{\pm}(\ell)\widehat{\mathcal{J}}_{j\beta 4}^{\pm}(\ell+k_{123}) \\
        &+\eta^{i\beta}\eta^{j\alpha}\mathcal{J}_{i\alpha 12}^{\pm}(\ell)\widehat{\mathcal{J}}_{j\beta 34}^{\pm}(\ell+k_{12})\\
        &+\eta^{i\beta}\eta^{j\alpha}\mathcal{J}_{i\alpha 23}^{\pm}(\ell)\widehat{\mathcal{J}}_{j\beta 14}^{\pm}(\ell+k_{23})
    \end{split}
\end{equation}
This one-loop integrand, we argue, is composed of just three distinct diagram types:~two $2$-gon, three $3$-gons, and one $4$-gon. Adopting the standard nomenclature, we will refer to these as the ``bubble'', ``triangle'', and ``box'' diagrams, respectively.   For simplicity, we begin with the second term (observe that the third contribution is exactly analogous):

\begin{widetext}
    \begin{align}
        \begin{split}
            \eta^{i\beta}\eta^{j\alpha}\mathcal{J}^{\pm}_{i\alpha 12}(\ell)\hat{\mathcal{J}}^{\pm}_{j\beta 34} (\ell +k_{12})= \frac{1}{\ell^2 (\ell + k_{12})^2} \bigg\{& 4\mathcal{A}^{\pm}_{12} \cdot \ell \ \mathcal{A}^{\pm}_{34} \cdot \ell + \eta^{i\beta} \mathcal{J}^{\pm}_{i\beta 1}(\ell) 
\mathcal{A}^{\pm}_{2} \cdot (\ell + k_1) \ \mathcal{A}^{\pm}_{34} \cdot  (\ell+k_{12})\bigg\}.
        \end{split}
    \end{align}
\end{widetext}

\begin{widetext}
    \begin{align}
    \label{eq:63}
        \begin{split}
            \eta^{i\beta}\eta^{j\alpha}\mathcal{J}^{\pm}_{i\alpha 12}(\ell)\hat{\mathcal{J}}^{\pm}_{j\beta 34} (\ell +k_{12})= \frac{1}{\ell^2 (\ell + k_{12})^2}\bigg\{& 4\mathcal{A}^{\pm}_{12} \cdot \ell \ \mathcal{A}^{\pm}_{34} \cdot \ell +\eta^{i\beta} \mathcal{J}^{\pm}_{i\beta 1}(\ell) 
\mathcal{A}^{\pm}_{2} \cdot (\ell + k_1) \ \mathcal{A}^{\pm}_{34} \cdot  (\ell+k_{12})\bigg\}
        \end{split}
    \end{align}
\end{widetext}
From this we can clearly identify one of the desired contributions. We are already familiar with this contribution, the ``bubble'' diagram, which appears as
\begin{equation}
    I^{\text{$2$-gon}}_{\ell}(1^{\pm}2^{\pm}\vert 3^{\pm}4^{\pm}) = \frac{4\mathcal{A}^{\pm}_{12} \cdot \ell \ \mathcal{A}^{\pm}_{34} \cdot \ell}{\ell^2 (\ell + k_{12})^2}.
\end{equation}
Now, to obtain the second contribution from (\ref{eq:63}), we use the recursion relation \eqref{eq:49} to express the factor $\eta^{i\beta} \mathcal{J}^{\pm}_{i\beta 1}(\ell)$ as
\begin{equation}
    \eta^{i\beta} \mathcal{J}^{\pm}_{i\beta 1}(\ell) = -\frac{4\mathcal{A}^{\pm}_1 \cdot \ell}{(\ell+k_1)^2},
\end{equation}
and from this we immediately have that the second contribution appearing in (\ref{eq:63}) is a $3$-gon or to a ``triangle'' diagram, and using the previously introduced notation, it is represented as $I^{\text{$3$-gon}}_{\ell}(1^{\pm}\vert 2^{\pm}\vert 3^{\pm} 4^{\pm})$. We can write it explicitly as
\begin{align}
    \begin{split}
    &I^{\text{$3$-gon}}_{\ell}(1^{\pm}\vert 2^{\pm}\vert 3^{\pm}4^{\pm})\\
    &\qquad  \qquad =- \frac{4 \varepsilon^{\pm}_{1} \cdot \ell  \ \varepsilon^{\pm}_{2} \cdot (\ell+k_{1}) \mathcal{A}^{\pm}_{34} \cdot (\ell+k_{12})  }{\ell^2(\ell + k_{1})^2(\ell + k_{12})^2}.
 \end{split}
\end{align}

We can observe that the third term in (\ref{eq:62}) is completely similar to the previous case, so it contributes with a $2$-gon $I^{\text{$2$-gon}}_{\ell}(2^{\pm}3^{\pm}\vert 4^{\pm}1^{\pm})$ and a $3$-gon $I^{\text{$3$-gon}}_{\ell}(2^{\pm}\vert3^{\pm}\vert 1^{\pm} 4^{\pm})$. Therefore, we only left to consider the first term in (\ref{eq:62}). By expanding it using the recursion relation, 

\begin{align}
\label{eq:67}
    \begin{split}
    &\eta^{i\beta}\eta^{j\alpha}\widetilde{\mathcal{J}}^{\pm}_{i\alpha 123}(\ell)\mathcal{J}^{\text{tp}\pm}_{j\beta 4} (\ell +k_{123})=\\
    &\frac{1}{\ell^2 (\ell + k_{123})^2}\bigg\{  \eta^{i\beta} \mathcal{J}^{\pm}_{i\beta 12}(\ell) 
\mathcal{A}^{\pm}_{3} \cdot (\ell + k_{12}) \ \mathcal{A}^{\pm}_{4} \cdot  (\ell+k_{123}) \\
&+\eta^{i\beta}\mathcal{J}^{\pm}_{i\beta 1}\mathcal{A}_{23}^{\pm}\cdot (\ell+k_1)\mathcal{A}_{4}^{\pm}\cdot (\ell+k_{123})\bigg\}
    \end{split}
\end{align}
We are already familiar with one of the contributions in (\ref{eq:67}), as it is just another $3$-gon,
\begin{equation}
    \begin{aligned}
       &I^{\text{$3$-gon}}_{\ell}(1^{\pm} \vert2^{\pm} 3^{\pm}\vert4^{\pm})\\
    &\qquad  \qquad =- \frac{4 \varepsilon^{\pm}_{1} \cdot \ell  \  \mathcal{A}^{\pm}_{23} \cdot (\ell+k_{1})\varepsilon^{\pm}_{4} \cdot (\ell+k_{123})  }{\ell^2(\ell + k_{1})^2(\ell + k_{123})^2}.
    \end{aligned}
\end{equation}

By expanding the only left over contribution, we find

\begin{widetext}
\begin{equation}
\label{eq:69}
\begin{aligned}
    \frac{1}{\ell^2 (\ell + k_{123})^2} \bigg\{&  \eta^{i\beta} \mathcal{J}^{\pm}_{i\beta 12}(\ell) 
\mathcal{A}^{\pm}_{3} \cdot (\ell + k_{12}) \ \mathcal{A}^{\pm}_{4} \cdot  (\ell+k_{123})\bigg\}=\\
&\frac{1}{\ell^2 (\ell + k_{123})^2} \bigg\{\frac{1}{(\ell+k_{12})^2}\bigg\{-4\mathcal{A}_{12}^{\pm}\cdot\ell \mathcal{A}^{\pm}_{3} \cdot (\ell + k_{12}) \ \mathcal{A}^{\pm}_{4} \cdot  (\ell+k_{123})\\
& +\frac{4}{(\ell+k_1)^2}\mathcal{A}_{1}^{\pm}\cdot\ell \mathcal{A}_{2}^{\pm}\cdot (\ell+k_1)\mathcal{A}^{\pm}_{3} \cdot (\ell + k_{12}) \ \mathcal{A}^{\pm}_{4} \cdot  (\ell+k_{123})\bigg\}\bigg\}
\end{aligned}
\end{equation}
\end{widetext}
We clearly observe two contributions here. One of them is another $3$-gon integrand that we may write down as

\begin{align}
    \begin{split}
 &I^{\text{$3$-gon}}_{\ell}(1^{\pm} 2^{\pm}\vert 3^{\pm}\vert4^{\pm})\\
    &\qquad  \qquad =- \frac{4 \mathcal{A}^{\pm}_{12} \cdot \ell\varepsilon^{\pm}_{3} \cdot (\ell+k_{12})  \ \varepsilon^{\pm}_{4} \cdot (\ell+k_{123})   }{\ell^2(\ell + k_{12})^2(\ell + k_{123})^2}.
 \end{split}
\end{align}

In the special case in which all four gluons have $+$ helicity, we may appeal to the first expression in \eqref{eq:25} to write the latter as

\begin{align}
    \begin{split}
    &I^{\text{$3$-gon}}_{\ell}(1^{+} \vert2^{+}\vert 3^{+}  4^{+}) \\
    &\:\:  =  \frac{4\langle r \vert \ell \vert k_{12}]\langle r \vert k_1 \vert k_{2}] \langle r \vert \ell + k_{12}\vert k_{3}] \langle r \vert \ell + k_{123} \vert k_{4}] }{\ell^2(\ell + k_{12})^2(\ell + k_{123})^2 \prod_{p=1}^{4}\langle r k_p \rangle }.
 \end{split}
\end{align}

The corresponding kinematic numerator for this ``triangle'' diagram, written accordingly as $N^{\text{$3$-gon}}_{\ell}(1^{+} 2^{+}\vert 3^{+} \vert 4^{+})$, is given by
\begin{align}
    \begin{split}
    &N^{\text{$3$-gon}}_{\ell}(1^{+} \vert2^{+}\vert 3^{+}  4^{+}) \\
    &\:\:  =  \frac{4\langle r \vert \ell \vert k_{12}]\langle r \vert k_1 \vert k_{2}] \langle r \vert \ell + k_{12}\vert k_{3}] \langle r \vert \ell + k_{123} \vert k_{4}] }{\prod_{p=1}^{4}\langle r k_p \rangle }.
 \end{split}
\end{align}

The second contribution in Eq. (\ref{eq:69}) corresponds to a box diagram represented as $I^{\text{$4$-gon}}_{\ell}(1^{\pm} \vert 2^{\pm}\vert 3^{\pm} \vert 4^{\pm})$ and defined by
\begin{align}
    \begin{split}
       & I^{\text{$4$-gon}}_{\ell}(1^{\pm} \vert 2^{\pm}\vert 3^{\pm} \vert 4^{\pm})\\
       &\qquad  = \frac{4\varepsilon^{\pm}_1 \cdot \ell \ \varepsilon^{\pm}_2 \cdot (\ell + k_1) \ \varepsilon^{\pm}_3 \cdot (\ell + k_{12}) \ \varepsilon^{\pm}_4 \cdot \ell}{\ell^2(\ell + k_1)^2 (\ell + k_{12})^2 (\ell + k_{123})^2}.
    \end{split}
\end{align}

If all four gluons have $+$ helicity, use of the the first expression in \eqref{eq:25} yields
\begin{align}
    \begin{split}
       & I^{\text{$4$-gon}}_{\ell}(1^{+} \vert 2^{+}\vert 3^{+} \vert 4^{+})\\
       &\:\:  = \frac{4\langle r\vert \ell + k_{123} \vert k_4] \langle r\vert \ell + k_{12} \vert k_3] \langle r\vert \ell + k_{1} \vert k_2] \langle r\vert \ell \vert k_1]}{\ell^2(\ell + k_1)^2 (\ell + k_{12})^2 (\ell + k_{123})^2 \prod_{p=1}^{4}\langle r k_p \rangle}.
    \end{split}
\end{align}
The corresponding kinematic numerator for this ``box'' diagram, denoted as $N^{\text{$4$-gon}}_{\ell}(1^{\pm} \vert 2^{\pm}\vert 3^{\pm} \vert 4^{\pm})$, takes the form
\begin{align}\label{eq:master4pt}
    \begin{split}
       & N^{\text{$4$-gon}}_{\ell}(1^{+} \vert 2^{+}\vert 3^{+} \vert 4^{+})\\
       &\:\:  = \frac{4\langle r\vert \ell + k_{123} \vert k_4] \langle r\vert \ell + k_{12} \vert k_3] \langle r\vert \ell + k_{1} \vert k_2] \langle r\vert \ell \vert k_1]}{\prod_{p=1}^{4}\langle r k_p \rangle}.
    \end{split}
\end{align}

As an aside, the following identity can be easily verified:
\begin{align}\label{eq:72}
    \begin{split}
        &N^{\text{$3$-gon}}_{\ell}(1^{+} 2^{+}\vert 3^{+} \vert 4^{+}) \\
        &\:\:  = N^{\text{$4$-gon}}_{\ell}(1^{+} \vert 2^{+}\vert 3^{+} \vert 4^{+}) - N^{\text{$4$-gon}}_{\ell}(2^{+} \vert 1^{+}\vert 3^{+} \vert 4^{+}).
    \end{split}
\end{align}
This aligns with the result derived in \cite{Boels:2013bi} using Feynman rules.

\subsection{The relation to full Yang-Mills theory}
It is well established in the literature that the self-dual sector of Yang-Mills theory provides valuable insights into the full Yang-Mills theory. This notable connection stems from two crucial aspects:~the self-dual sector is a subset of the full theory, and secondly, both share identical one-loop scattering amplitudes under specific gluon helicity configurations, namely when all gluons have $+$ helicity, or when the first gluon has $-$ helicity and the rest have $+$ helicity. A comprehensive analysis of the full Yang-Mills theory, using the approach we have developed in this paper, was previously conducted in \cite{Gomez:2022dzk}. Rather than repeating that entire analysis, we will focus on the recursion relations for one-loop gluon preintegrands in the full theory. These relations are given by
\begin{widetext}
\begin{align}\label{eq:76}
\begin{split}
 \mathcal{J}_{P \mu \nu}(\ell)&= \frac{1}{(\ell+k_P)^2} \bigg\{  \mathcal{A}_{P \rho}(\delta_\mu^\rho\left(\ell+2k_P\right)_\nu+\delta_\nu^\rho\left(\ell-k_P\right)_\mu-\eta_{\mu \nu}\left(2 \ell+k_{ P}\right)^\rho) \\
 &\quad \: +\sum_{P=Q R}\mathcal{A}_{Q \rho} \mathcal{A}_{R \sigma} (2 \delta_\mu^\rho \delta_\nu^\sigma-\delta_\mu^\sigma \delta_\nu^\rho-\eta_{\mu \nu} \eta^{\rho \sigma}) \\
&\quad \: +\sum_{P=Q R}\mathcal{J}_{Q \rho \nu}(\ell) \mathcal{A}_{R \sigma}(\delta_\mu^\sigma\left(\ell + k_{P}+k_R\right)^\rho-\delta_\mu^\rho\left(2 \ell + k_{ P}+k_{ Q}\right)^\sigma+\eta^{\rho \sigma}\left(\ell +k_Q-k_R\right)_\mu)  \\
&\quad \: +(2 \delta_\gamma^\rho \delta_\mu^\sigma-\eta^{\rho \sigma} \eta_{\mu \gamma}-\delta_\gamma^\sigma \delta_\mu^\rho) \sum_{P=Q R S} \mathcal{J}_{Q \rho \nu}(\ell) \mathcal{A}_{R \sigma} \mathcal{A}_S^\gamma \bigg\}.
\end{split}
\end{align}
\end{widetext}
When compared to the corresponding one-loop preintegrand in the self-dual sector \eqref{eq:49}, the full sector's preintegrand is notably more intricate and consists of a considerably greater number of terms. However, starting from this, it is possible to verify that the self-dual sector can be considered, metaphorically speaking, as a ``subset'' of the full sector. To illustrate this point, we will define a subrecursion based on \eqref{eq:76} for a ``truncated'' preintegrand $\mathcal{J}'_{P\mu\nu}(\ell)$, characterized by
    \begin{eqnarray}
    \mathcal{J}'_{P\mu\nu}(\ell)&=&  \frac{1}{(\ell+k_P)^2} \bigg\{ \mathcal{A}_{P\rho}(\delta^{\rho}_{\mu}k_{P\nu}-\eta_{\mu\nu}\ell^{\rho} )\\
    &&+\sum_{P=QR}\mathcal{J}'_{Q\rho\nu}(\ell)\mathcal{A}_{R\sigma}(\delta^{\sigma}_{\mu}k_{R}^{\rho}-\delta^{\rho}_{\mu}(\ell + k_Q)^{\sigma} ) \bigg\}.\notag
\end{eqnarray}
Following this subrecursion, we can then define, as before, the on-shell one-loop integrand for each deconcatenation $P=Q\vert R$ with the corresponding restricted cyclic completion $(QR)$ for the truncated pre-integrand given by
\begin{eqnarray}
    I^{\text{$1$-loop}}_{Pn}(\ell) &=&\eta^{i\beta}\eta^{j\alpha}\widetilde{\mathcal{J}}_{i\alpha P}^{'}(\ell)\hat{\mathcal{J}}_{j\beta n}^{'}(\ell+k_{P})\\
    &&+ \sum_{P=(Q\,Rn)}\eta^{i\beta}\eta^{j\alpha}\mathcal{J}_{i\alpha Q}^{'}(\ell)\hat{\mathcal{J}}_{j\beta Rn}^{'}(\ell+k_{Q})\notag
\end{eqnarray}
subject to momentum conservation $k_P=0$ and the on-shell transversality condition $\varepsilon_p \cdot k_p = 0$ for every single index Berends-Gielle current. By making the corresponding substitutions, we obtain the explicit expression
\begin{widetext}
    \begin{align}
        \begin{split}
            \eta^{i\beta}\eta^{j\alpha}\mathcal{J}_{i\alpha Q}^{'}(\ell)\hat{\mathcal{J}}_{j\beta R}^{'}(\ell+k_{Q})=\frac{1}{\ell^2(\ell+k_Q)^2}\bigg\{& 4\mathcal{A}_{Q}\cdot\ell \ \mathcal{A}_{R}\cdot \ell  +\sum_{Q=ST}\eta_{\mu\nu} \mathcal{J}^{\prime\mu\nu}_{S}(\ell)  \mathcal{A}_{T}\cdot (\ell+k_{S}) \ \mathcal{A}_{R}\cdot (\ell + k_Q)\bigg\}
        \end{split}
    \end{align}
\end{widetext}

After fixing the helicities and converting to light-cone coordinates, this yields exactly the same contributions as those in equation \eqref{eq:57}. A similar situation is observed in the one-loop pre-integrands for the ghost fields with the restricted cyclic completion $(QR)$
\begin{equation}
 \begin{aligned}
        \widetilde{B}_P(\ell) &= \frac{1}{\ell^2}  \sum_{P=(QR)} B_Q(\ell) \ \mathcal{A}_R \cdot (\ell + k_Q), \\
         \widetilde{C}_P(\ell) &= \frac{1}{\ell ^2}  \sum_{P=(QR)} C_Q(\ell) \ \mathcal{A}_R \cdot (\ell + k_P),
\end{aligned}   
\end{equation}
as it is not difficult to see that, for the case $P=1234$, one recovers bubble, triangle and box diagrams, so we just recovered the (anti)-self-dual Yang-Mills integrands.

\section{Multiparticle solutions of self-dual and anti-self-dual gravity}\label{section:gr-dc}
In this section, we focus on the multiparticle solutions to equations \eqref{eq:14}, which will yield multi-particle solutions for the self-dual and anti-self-dual equations \eqref{eq:12}. Similar to the approach taken in the previous section, these solutions are obtained through a perturbiner expansion, resulting in the recursive generation of fully off-shell Berends-Giele currents. We will demonstrate that from these currents, we can extract numerators that can be expressed as the ``square'' of the kinematic numerators derived from the off-shell color-dressed Berends-Giele currents for the self-dual and anti-self-dual sectors of Yang-Mills theory. Furthermore, by applying the modified version of the sewing procedure from \cite{Gomez:2022dzk} to recursively produce one-loop off-shell currents, we will also be able to extract numerators that exhibit the same property of being expressible as the ``square'' of the pre-integrand numerators for these sectors.

\subsection{Perturbiner}
We proceed first to write the multiparticle solutions in the form of a perturbiner expansion. Thus, we propose solutions ansätze of the equations in \eqref{eq:14} of the form
\begin{equation}
    \psi^{\pm}(x) = \sum_{n \geq 1} \sum_{P \in \mathcal{OW}_n} \Psi^{\pm}_P \ue^{ik_P \cdot x}.
\end{equation}
Inserting this back into said equations yields the recursion relations
\begin{equation}\label{eq:83X}
    \Psi^{\pm}_P = \pm \frac{1}{2s_P} \sum_{P = Q \cup R} \Pi^{\pm ij} k_{Qi}k_{Rj}\Pi^{\pm kl} k_{Qk}k_{Rl}\Psi_{Q}^{\pm}\Psi_{R}^{\pm}.
\end{equation}
Now, our attention is primarily on the perturbiner expansions for the self-dual and anti-self-dual graviton fields $h^{\pm}_{\mu\nu}$. In the foregoing notation, these assume the form
\begin{equation}
    h^{\pm}_{\mu\nu}(x) = \sum_{n \geq 1} \sum_{P \in \mathcal{OW}_n} H^{\pm}_{\mu\nu P} \ue^{ik_P \cdot x},
\end{equation}
where in compliance with the gauge-fixing constraints and the relation specified in equation \eqref{eq:13}, we must have $H^+_{\mu u P} = H^+_{\mu \bar{z} P} = 0$, $H^-_{\mu u P} = H^-_{\mu z P} = 0$ and
\begin{equation}\label{eq:85X}
    H^{\pm}_{\alpha\beta P} = -\Pi_{\alpha}^{\pm i}\Pi_{\beta}^{\pm j} k_{P i} k_{P j} \Psi^{\pm}_{P}.
\end{equation}
Applying \eqref{eq:83X} to \eqref{eq:85X}, we can, with some algebraic work, transform the latter into
\begin{widetext}
\begin{equation}\label{eq:86X}
    \begin{aligned}
        H^{\pm}_{\alpha\beta P} = \mp \frac{1}{2s_P} \sum_{P = Q \cup R} \{ k_{Q}^{\gamma} k_{Q}^{\lambda} H^{\pm}_{\gamma\lambda R} H^{\pm}_{\alpha\beta Q} + k_{R}^{\gamma} k_{R}^{\lambda} H^{\pm}_{\gamma\lambda Q} H^{\pm}_{\alpha\beta R} -  k_{R}^{\gamma}  H^{\pm}_{\alpha \gamma Q} k_{Q}^{\lambda}H^{\pm}_{\lambda \beta R} - k_{Q}^{\gamma}  H^{\pm}_{\alpha \gamma R} 
k_{R}^{\lambda}H^{\pm}_{\lambda \beta Q} \}.
    \end{aligned}
\end{equation}
\end{widetext}
Relaxing the on-shell requirement for single-particle states enables us to interpret $H^{\pm}_{\alpha\beta P}$ as off-shell Berends-Giele currents for the self-dual and anti-self-dual graviton fields $h^{\pm}_{\mu\nu}$.

We should also examine the amplitudes corresponding to the self-dual and anti-self-dual sectors in gravity and their connection to off-shell Berends-Giele currents. For this, we follow the exact same approach as before. Specifically, we decompose the single index Berends-Giele currents into their kinematic degrees of freedom according to $H^{\pm}_{\alpha\beta p} = \bar{\varepsilon}^{\pm}_{\alpha p} \varepsilon^{\pm}_{\beta p}$, and then the amplitude for the scattering of $n$ gravitons can be represented as
\begin{align}
    \begin{split}
        &\mathcal{M}^{\mathrm{tree}}_n(1^{h_1},2^{h_1},\dots,(n-1)^{h_1},n^{h_n}) \\
        &\qquad \qquad  = s_{12 \cdots (n-1)} \eta^{i\alpha}\eta^{i\beta}H^{h_1 a_n}_{\alpha \beta 12 \cdots (n-1)} \bar{\varepsilon}^{h_n}_{i n}\varepsilon^{h_n}_{j n},
    \end{split}
\end{align}
where, as usual, momentum conservation is assumed. It is important to note that the gauge-fixing constraints lead to $\bar{\varepsilon}^{\pm}_{i n}\varepsilon^{\pm}_{j n} = 0$, regardless of how the light-cone coordinates are grouped. This results in
\begin{equation}
    \mathcal{M}^{\mathrm{tree}}_n(1^{\pm},2^{\pm},\dots,(n-1)^{\pm},n^{\pm}) = 0.
\end{equation}
This means that only tree-level amplitudes of the form $\mathcal{M}^{\mathrm{tree}}_n(1^{\pm},2^{\pm},\dots,(n-1)^{\pm},n^{\mp})$ could potentially be non-zero. We will demonstrate that these tree-level amplitudes actually vanish for $n > 3$, but before that, we need to examine how the double copy prescription can be articulated in terms of perturbiners.

\subsection{Double copy relations for Berends-Giele currents}\label{subsection:DC}
In this short section, we aim to outline how numerators can be extracted from the off-shell Berends-Giele currents \eqref{eq:86X} that can be expressed as the ``square'' of the kinematic numerators obtained from the off-shell color-dressed Berends-Giele currents for the self-dual and alti-self-dual Yang-Mills fields. We will follow the approach presented in \S10 of \cite{Escudero:2022zdz}.

The key point to consider is that in the decomposition of the kinematic degrees of freedom of single index Berends-Giele currents, both $\bar{\varepsilon}^{\pm}_{\alpha p}$ and $\varepsilon^{\pm}_{\alpha p}$ should be regarded as generators of two infinite-dimensional Lie algebras $\bar{\mathfrak{g}}^{\prime\pm}$ and $\mathfrak{g}^{\prime\pm}$, whose Lie bracket is defined in both cases by the expression \eqref{eq:41}. Starting from this premise and using the notation introduced in section \ref{sec:IIIC}, one can show that the recursion relation \eqref{eq:86X} can be reformulated as
\begin{equation}
     H^{\pm}_{\alpha\beta P} = \left( \pm \tfrac{1}{2} \right)^{\lvert P \rvert - 1} \llbracket \bar{\varepsilon}^{\pm}_{\alpha} \otimes \varepsilon^{\pm}_{\beta} \rrbracket \circ  b_{\mathrm{cd}}(P).
\end{equation}
We can compare this with the expression for the color-dressed Berends-Giele current given by \eqref{eq:46}. The resemblance between these expressions leads us to conclude that, aside from the $\frac{1}{2}$ factor, they are identical if we substitute $c^{a}$ with $\bar{\varepsilon}^{\pm}_{\alpha}$. This is what we mean when we say that the numerators of the Berends-Giele current $H^{\pm}_{\alpha\beta P}$ can be built as the “square” of the kinematic numerators of the color-dressed Berends-Giele current $\mathcal{A}^{\pm a}_{\alpha P}$.

\subsection{The zeroth copy of self-dual and anti-self-dual Yang-Mills}
In addition to the double copy relations at the level of Berends-Giele currents, we must also address the zero copy in the self-dual and anti-self-dual sectors of Yang-Mills theory. For this we will draw from the approach presented in \S11 and \S12 of \cite{Escudero:2022zdz}.

To define the zero copy, we introduce a second Lie algebra $\bar{\mathfrak{g}}$ associated with a compact Lie group. We select generators $\bar{T}^{\bar{a}}$ for this Lie algebra and denote its structure constants as $\bar{f}^{\bar{a}\bar{b}\bar{c}}$. The fundamental dynamic variable will be a function $\xi^{\pm}$ defined on $\mathbb{R}^{1,3}$ taking values in $\mathfrak{g} \otimes \bar{\mathfrak{g}}$,  where the sign $\pm$ indicates whether we are dealing with a self-dual or anti-self-dual Yang-Mills field. This function can, of course, be expressed as $\xi^{\pm} = \xi^{\pm a \bar{a}}T^{a} \otimes \bar{T}^{\bar{a}}$. In terms of this notation, we proceed to define
\begin{equation}
 \{\!\!\{ \xi^{\pm},\xi^{\pm} \}\!\!\} = f^{abc} \bar{f}^{\bar{a}\bar{b}\bar{c}} \xi^{b\bar{b}} \xi^{\pm c\bar{c}} T^{a} \otimes \bar{T}^{\bar{a}}.
\end{equation}
The dynamics of the theory is then governed by the equation of motion
\begin{equation} \label{eq:91X}
\Box \xi^{\pm} \pm \tfrac{1}{2} \{\!\!\{ \xi^{\pm},\xi^{\pm} \}\!\!\} = 0,
\end{equation}
Note that, effectively, what has been done is to replace in equation \eqref{eq:9} the effect of the antisymmetric tensor $i \Pi^{\pm ij}$ with that of the Lie bracket of $\bar{\mathfrak{g}}$.

We are now interested in multiparticle solutions of \eqref{eq:91X}. These come in two flavours, analogous to those of equation \eqref{eq:9}:~color-stripped and color-dressed perturbiners. For the color-stripped perturbiner, we propose an solution ansatz of the form
\begin{equation}
    \xi^{\pm}(x) = \sum_{n \geq 1} \sum_{P,Q \in \mathcal{W}_n} \Xi^{\pm}_{P \vert Q}  \ue^{ik_P \cdot x} T^{a_P} \otimes \bar{T}^{\bar{a}_P}.
\end{equation}
Upon inserting this into equation \eqref{eq:91X}, we obtain the recursion relations
\begin{equation}\label{eq:93X}
   \Xi^{\pm}_{P \vert Q} = \pm \frac{1}{s_P} \sum_{P = RS} \sum_{Q=TU} (\Xi^{\pm}_{R \vert T}\Xi^{\pm}_{S \vert U} - \Xi^{\pm}_{S \vert T} \Xi^{\pm}_{R \vert U}).
\end{equation}
 Notice that, due to the antisymmetry of the right-hand side of \eqref{eq:93X} under the exchange of words $R$ and $S$, and $T$ and $U$, these coefficients automatically satisfy the shuffle constraint $\Xi^{\pm}_{P \shuffle Q \vert R} = 0$. This property justifies the use of the term Berends-Giele double currents to refer to $\Xi^{\pm}_{P \vert Q}$. As for the color-dressed perturbiner, the proposed solution ansatz is
 \begin{equation}
     \xi^{\pm a\bar{a}}(x) = \sum_{n \geq 1} \sum_{P,Q \in \mathcal{OW}_n} \Xi^{\pm a \bar{a}}_{P}  \ue^{ik_P \cdot x}.
 \end{equation}
The substitution of this into the componentwise representation of equation \eqref{eq:91X} leads to the recursion relations
 \begin{equation}
     \Xi^{\pm a \bar{a}}_{P} = \pm \frac{1}{s_P}  \sum_{P = Q \cup R} f^{abc} \bar{f}^{\bar{a}\bar{b}\bar{c}} \Xi^{\pm b \bar{b}}_{Q} \Xi^{\pm c \bar{c}}_{R}.
 \end{equation}

 In order to find a similar factorization as of color-kinematics and double-copy, we introduce the notation
 \begin{equation*}
     c_{L[P]}^{a}=f_{a_{p_{1}}a_{p_2}}^{\;\;b}f_{ba_{p_3}}^{\;\;c}\cdots f_{da_{p_{n-1}}}^{\;\;e}f_{ea_{p_n}}^{\;\;a},
 \end{equation*}
 and similarly for $\bar{c}^{\bar{a}}_{L[P]}$, with the expected conventions $c^{a}_{p}=\delta^{a}_{a_{p_n}}$ and $\bar{c}^{\bar{a}}_{p}=\delta^{\bar{a}}_{\bar{a}_{p_n}}$. Therefore, it is now straightforward to show that tree-level self-dual Yang-Mills zeroth-copy is written in terms of the color-dressed Berends-Giele map as 
\begin{equation}
    \Xi_{P}^{\pm a\bar{a}}=(\pm 1)^{\lvert P \rvert -1} \llbracket c^{a}\otimes \bar{c}^{\bar{a}}\rrbracket \circ b_{\text{cd}}(P),
\end{equation}
that is, equivalent to replacing $\varepsilon_{\mu}^{\pm}$ by $\bar{c}^{\bar{a}}$ from the color-kinematics factorization.

\subsection{Tree-level KLT relations in self-dual and anti-self-dual Yang-Mills}

In \ref{subsection:DC} it was observed that tree-level self-dual gravity scattering amplitudes can be expressed in terms of self-dual Yang-Mills colored-amplitudes via the double copy of the Berends-giele currents. Here, Following \cite{Frost_2023}, we find the standard KLT formulation and show that the KLT kernel is equal to the one form Yang-Mills.  consider the KLT map $S:\mathcal{L}\to \mathcal{L}^{*}$, where $\mathcal{L}$ denotes the space of Lie polynomials and $\mathcal{L}^{*}$ its dual, defined for a Lie monomial $\Gamma$ by $S(\Gamma)=\{\Gamma \}$.
The $S$ map is nothing but the inverse of the Berends-Giele map $b_{cs}$, and has the associated momentum kernel $S(P|Q)_{p}$, labeled by permutations $P$ and $Q$ of $12\cdots\hat{p}\cdots(n-1)$. These matrix elements, as it was shown, work as inverses for the Berends-Giele double currents $\Xi_{pP|pQ}^{\pm}$
\begin{equation}
\sum_{R}\Xi_{pP|pR}^{\pm}S(R|Q)_{p}=\sum_{R}S(P|R)_{p}\Xi_{pR|pQ}^{\pm}=\delta_{P,Q}.
\end{equation}
The standard KLT matrix is recursively computed by 
\begin{equation}
\begin{split}
    S(\varnothing\mid\varnothing)_{p}&=1,\\
    S(Pq\mid QqR)_{p}&=k_{q}\cdot k_{pQ}S(P\mid QR).
\end{split}
\end{equation}
Just as the double currents can be used to explicitly compute the Berends-Giele currents under the prescription
\begin{equation}
    \mathcal{A}_{\alpha1P}^{\pm}=\sum_{Q}\Xi_{1P\mid 1Q}^{\pm}\varepsilon_{\alpha L[1Q]}^{\pm},
\end{equation}
the momentum kernel, as its inverse, is used to compute the master numerators,
\begin{equation}
\label{98}
    \varepsilon_{\alpha L[1Q]}^{\pm}=\sum_{P}S(P\mid Q)_{1}\mathcal{A}_{\alpha1P}^{\pm}.
\end{equation}
Using the fact that the self-dual gravity currents can be expressed in terms of a master numerator $\bar{\varepsilon}_{L[P]\alpha}$ and the self-dual Yang-Mills currents as
\begin{equation}
    H_{1\cdots(n-1)\alpha\beta}^{\pm}=\sum_{P}\bar{\varepsilon}_{L[1P]\alpha}^{\pm}\mathcal{A}_{1P\beta}^{\pm},
\end{equation}
we use Eq. (\ref{98}) to obtain a KLT-type relation for the currents
\begin{equation}
    \label{80}
    H_{1\cdots(n-1)\alpha\beta}^{\pm}=\sum_{P,Q}\bar{\mathcal{A}}_{1P\alpha}^{\pm}S(P\mid Q)_{1}\mathcal{A}_{1Q\beta}^{\pm}.
\end{equation}
From this KLT-type relation there is just one step to find a relation for the amplitudes, as we just need to use the prescription we have to calculate the $n$-point amplitude from the $(n-1)$-points Berends-Giele current
\begin{equation}
\label{81}
    \mathcal{M}_{n}^{\text{tree}}=\sum_{P,Q}\frac{1}{s_{1P}}\bar{A}^{\text{tree}}(1Pn)S(P\mid Q)_{1}A^{\text{tree}}(1Qn),
\end{equation}
obtaining the so-called $(n-2)!$ version of KLT relation for the tree-level amplitudes. As is known, these are equivalent to the more common $(n-3)!$ KLT relations on-shell \cite{Bjerrum_Bohr_2010}. At this point we should state some important remarks. As it is expected, the standard matrix elements $S(P\mid Q)_{1}$ in the self-dual sector are equal to the full Yang-Mills KLT matrix elements.

\subsection{One-loop self-dual gravity}
We now wish to understand the one-loop integrand for self-dual gravity in the perspective of the off-shell perturbiner expansion. The pre-integrand for self-dual gravity is similarly obtained as in self-dual and anti-self-dual Yan-Mills. We consider a Berends-Giele current $H_{\alpha\beta lP}^{\pm}$, and by factorizing explicitly two-copies of single-particle current for the loop-closing leg $H_{lij}^{\pm}=h_{lij}^{\pm}$, we find an expression for the gravity pre-integrand
\begin{equation}
H_{\alpha\beta}^{\pm}=h_{l}^{ij\pm}\mathcal{K}_{ij\alpha\beta}^{\pm}(\ell)
\end{equation}
with a direct computation resulting in
\begin{widetext}
\begin{equation}
\begin{split}
    \mathcal{K}_{ij\alpha\beta P}^{\pm}(\ell)=&\frac{1}{2(\ell + k_P)^2 }\bigg\{H_{\sigma_{1}\sigma_{2}P}^{\pm}\eta_{i\rho_{1}}\eta_{j\rho_{2}}\bigg\{\ell^{\sigma_{1}}\ell^{\sigma_{2}}\delta^{\rho_{1}}_{\alpha}\delta^{\rho_{2}}_{\beta} -\ell^{\sigma_{1}}k^{\rho_{2}}_{P}\delta^{\rho_{1}}_{\alpha}\delta^{\sigma_{2}}_{\beta} -k^{\sigma_{1}}_{P}\ell^{\rho_{2}}\delta^{\rho_{1}}_{\alpha}\delta^{\sigma_{2}}_{\beta} +  k^{\sigma_{1}}_{P }k^{\sigma_{2}}_{P }\delta^{\rho_{1}}_{\alpha}\delta^{\rho_{2}}_{\beta} \bigg\} \\
    & + \sum_{P=Q\cup R}H_{\sigma_{1}\sigma_{2}R}^{\pm}\mathcal{K}_{ij\rho_{1}\rho_{2}Q}^{\pm}\bigg( (\ell + k_{Q})^{\sigma_{1}}(\ell + k_{Q})^{\sigma_{2}}\delta^{\rho_{1}}_{\alpha}\delta^{\rho_{2}}_{\beta} - (\ell+k_Q)^{\sigma_{1}}k^{\rho_{2}}_{R}\delta^{\rho_{1}}_{\alpha}\delta^{\sigma_{2}}_{\beta} \\
    &- k^{\sigma_{1}}_{R}(\ell+k_{R})^{\rho_{2}}\delta^{\rho_{1}}_{\alpha}\delta^{\sigma_{2}}_{\beta} + k^{\sigma_{1}}_{R}k^{\sigma_{2}}_{R}\delta^{\rho_{1}}_{\alpha}\delta^{\rho_{2}}_{\beta} \bigg) \bigg\}
\end{split}
\end{equation}   
\end{widetext}

The prescription that generates all the contributions to the one-loop integrand for self-dual gravity without tadpoles and without redundant contributions is described in \cite{gomez2024}. Applied to our set-up in self-dual Gravity, we have to modify the current to get rid off the overcounting of equivalent diagrams. The modified current is as follows: 

\begin{widetext}
\begin{equation}
\begin{split}
    \tilde{\mathcal{K}}_{ij\alpha\beta P}^{\pm}(\ell)=&\frac{1}{2(\ell + k_P)^2 }\bigg\{f_1(\vert P\vert )H_{\sigma_{1}\sigma_{2}P}^{\pm}\eta_{i\rho_{1}}\eta_{j\rho_{2}}\bigg\{\ell^{\sigma_{1}}\ell^{\sigma_{2}}\delta^{\rho_{1}}_{\alpha}\delta^{\rho_{2}}_{\beta} -\ell^{\sigma_{1}}k^{\rho_{2}}_{P}\delta^{\rho_{1}}_{\alpha}\delta^{\sigma_{2}}_{\beta} -k^{\sigma_{1}}_{P}\ell^{\rho_{2}}\delta^{\rho_{1}}_{\alpha}\delta^{\sigma_{2}}_{\beta} +  k^{\sigma_{1}}_{P }k^{\sigma_{2}}_{P }\delta^{\rho_{1}}_{\alpha}\delta^{\rho_{2}}_{\beta} \bigg\} \\
    & + \sum_{P=Q\cup R}g_{n}H_{\sigma_{1}\sigma_{2}R}^{\pm}g_(\vert Q\vert)\tilde{\mathcal{K}}_{ij\rho_{1}\rho_{2}Q}^{\pm}\bigg( (\ell + k_{Q})^{\sigma_{1}}(\ell + k_{Q})^{\sigma_{2}}\delta^{\rho_{1}}_{\alpha}\delta^{\rho_{2}}_{\beta} - (\ell+k_Q)^{\sigma_{1}}k^{\rho_{2}}_{R}\delta^{\rho_{1}}_{\alpha}\delta^{\sigma_{2}}_{\beta} \\
    &- k^{\sigma_{1}}_{R}(\ell+k_{R})^{\rho_{2}}\delta^{\rho_{1}}_{\alpha}\delta^{\sigma_{2}}_{\beta} + k^{\sigma_{1}}_{R}k^{\sigma_{2}}_{R}\delta^{\rho_{1}}_{\alpha}\delta^{\rho_{2}}_{\beta} \bigg) \bigg\}
\end{split}
\end{equation}  
\end{widetext}
where the coefficients $f_{1}$ and $g$ are included here to balance the overcounting. As it is shown in \cite{gomez2024},
\begin{equation}
    f_1=\frac{1}{2},\qquad g(\vert q\vert)=\frac{\vert Q\vert}{\vert P\vert}
\end{equation}
In order to define the on-shell one-loop graviton integral, we have to make further simplifications; that implies that we have to remove the tadpole and impose a condition in the length of the deconcatenations as $\vert R\vert >1$ so we avoid external leg bubbles. Therefore, the one-loop integrand for graviton in the self-dual sector is given by
\begin{equation}
\label{eq:106}
    \bar{I}^{\text{$1$-loop}}_{P}(\ell) = \widetilde{\mathcal{K}}_{P\alpha\beta }^{\pm\alpha\beta}(\ell)
\end{equation}

Once again, when going to the on-shell integrand, the momentum conservation $k_P = 0$ and the on-shell transversality condition $\varepsilon^{\pm}_p \cdot k_p = 0$ for every single index Berends-Giele current are understood. Similar considerations are taking when analyzing the on-shell contributions as in eq. \eqref{eq:54}, where the transversality condition for single index Berends-Giele currents where used to simplify significantly the computations. Thus, by implementing the above prescription, we observe the on-shell integrand has the simple form

    \begin{equation}
    \label{eq:107}
        \begin{split}
          &I^{\text{$1$-loop}}_{P}(\ell)=\frac{1}{\ell^2 (\ell + k_Q)^2}\times\\
          &\quad\quad\sum_{\mathclap{\substack{P=Q\cup R\\ \vert R\vert >1}}}H_{\sigma_{1}\sigma_{2}R}^{\pm}g(\vert Q\vert)\tilde{\mathcal{K}}_{\alpha\beta Q}^{\pm\alpha\beta} (\ell + k_{Q})^{\sigma_{1}}(\ell + k_{Q})^{\sigma_{2}}
        \end{split}
    \end{equation}

Here we are applying the same idea that of self-dual Yang-Mills, as some of the leftover contractions in \eqref{eq:106} will be zero on-shell, so we can ignore them and reduce the integral to the meaningful terms. Applying further deconcatenations, we can check once again that after $k-1$ deconcatentions of $P$ we obtain $k$ subwords $P_{1},\dots, P_{k}$ which involve contractions of Berends-Giele currents $H^{\pm}_{\alpha\beta}$ with loop momenta possibly shifted by $k_{P_{i}}+\cdots+k_{P_{j}}$ for $i<j$. Using the terminology of Section D, we refer these contributions as $I^{\text{$k$-gon}}_{\ell}(P_{1}^{\pm}\vert\cdots\vert P_{k}^{\pm})$.
Clearly, we have already encountered the $2$-gon contribution for a deconcatenation of the word $P$ in  word $QR$ as
\begin{equation}
\begin{split}
    I^{\text{$2$-gon}}_{\ell}(Q^{\pm}\vert R^{\pm})=\frac{4g(\vert Q\vert)H^{\pm}_{\alpha\beta Q} \ell^{\alpha}\ell^{\beta}\ H^{\pm}_{\eta\gamma R} \ell^{\eta}\ell^{\gamma}}{\ell^2 (\ell + k_Q)^2}    
\end{split}
\end{equation}
Observe that we get the same $4$ pre-multiplying the $2$-gon as in \eqref{eq:58}. In general, we are able to write the total contribution for a word $P$ as a linear combination of all the possible $k$-gons,
\begin{equation}
I^{\text{$1$-loop}}_P(\ell) = \sum_{k=2}^{n} \sum_{P=P_1 \cdots P_k} I^{\text{$k$-gon}}_{\ell}(P^{\pm}_1 \vert \cdots \vert P^{\pm}_k).
\end{equation}

We can again illustrate this construction for the word $P=1234$. Following the prescription in (\ref{eq:107}), we have to consider 
\begin{equation}
\label{eq:110}
        \begin{split}
           I^{\text{$1$-loop}}_{1234}(\ell)&=\frac{1}{\ell^2 (\ell + k_{123})^2} H_{\sigma_{1}\sigma_{2}34}^{\pm}g(\vert 12\vert)\tilde{\mathcal{K}}_{\alpha\beta 12}^{\pm\alpha\beta} \\
           &\quad\times(\ell + k_{12})^{\sigma_{1}}(\ell + k_{12})^{\sigma_{2}} +(\text{permts})
        \end{split}
\end{equation}
These steps are similar as those described for self-dual Yang-Mills, so we may go on an write down the contribution we encounter from the second term as the $2$-gon or the bubble diagram
\begin{equation}
\begin{aligned}
    &I^{\text{$2$-gon}}_{\ell}(1^{\pm}2^{\pm}\vert 3^{\pm}4^{\pm})=\\
    &\frac{2 H_{\mu\nu34}(\ell+ k_{12})^{\mu}(\ell + k_{12})^{\nu }} {\ell^2}\frac{H_{\alpha\beta12}\ell^{\alpha}\ell^{\beta} }{(\ell+k_{12})^2}   
\end{aligned}
\end{equation}
Now, from the first term in \eqref{eq:110} we encounter two contributions, which represents the triangle and the box diagrams. The former one is the $3$-gon, given by
\begin{equation}
        \begin{aligned}
            &I^{\text{$3$-gon}}_{\ell}(1^{\pm}2^{\pm}|3^{\pm}|4^{\pm})=
            \quad\frac{2\bar{\varepsilon}_{4}\cdot \ell\varepsilon_{4}\cdot \ell }{\ell^2}\\
            &\quad\quad\frac{\bar{\varepsilon}_{3}\cdot (\ell +k_{12})
            \varepsilon_{3}\cdot (\ell+ k_{12}) }{(\ell +k_{123})^2}\frac{H^{\pm}_{\mu\nu12}(\ell)^{\mu}(\ell)^{\nu}}{(\ell+k_{12})^2}.
        \end{aligned}
    \end{equation}
     In the case of fixing positive helicities, we may use spinor-helicity formalism (\ref{eq:25}) to write down $3$-gon integrands in the following way 
 \begin{equation}
 \begin{aligned}
    &I^{\text{$3$-gon}}_{\ell}(1^{+}2^{+}|3^{+}|4^{+})=\frac{2}{\ell^{2}(\ell+k_{12})^{2}(\ell-k_{4})^{2}k_{12}^2}\\
    &\times\left(\frac{\langle q|1|2]\langle q|\ell|1+2]\langle q|\ell+1+2|3]\langle q|\ell-4|4]}{\prod_{i}^{4}\langle qi\rangle}\right)^2
 \end{aligned}
\end{equation}
The expression for the $4$-gon is similarly found by taking further deconcatenations,
\begin{eqnarray}
         &&I^{\text{$4$-gon}}_{\ell}(1^{\pm}|2^{\pm}|3^{\pm}|4^{\pm})=\notag\\
         &&\quad\frac{2\bar{\varepsilon}_{4}\cdot\ell\varepsilon_{4}\cdot\ell}{\ell^2}\frac{\bar{\varepsilon}_{3}\cdot (\ell+k_{12})\varepsilon_{3}\cdot (\ell+k_{12})}{(\ell +k_{123})^2}\\
         &&\quad\frac{\bar{\varepsilon}_{2}\cdot (\ell+k_1)\varepsilon_{2}\cdot (\ell+k_1)}{(\ell+k_{12})^2}\frac{\bar{\varepsilon}_{1}\cdot \ell\varepsilon_{1}\cdot \ell}{(\ell+k_1)^2}.\notag 
\end{eqnarray}

Then, by fixing positive helicities, we can write it in the standard form 
    \begin{equation}
 \begin{aligned}
     &I^{\text{$4$-gon}}_{\ell}(1^{+}|2^{+}|3^{+}|4^{+})=\frac{2}{\ell^2(\ell+k_{1})^{2}(\ell+k_{12})^{2}(\ell-k_{4})^{2}}\\
     &\times\left(\frac{\langle q| \ell-4 |4 ] \langle q| \ell+1+2|3]\langle q|\ell+1|2]\langle q|\ell|1]}{\prod_{i}^{4}\langle qi\rangle}\right)^{2}         
 \end{aligned}
 \end{equation}

Now, with the master numerators we have found in \eqref{eq:master4pt}, we apply the double copy procedure to obtain the self-dual gravity numerators as the square of the self-dual Yang-Mills ones. More explicitly, we have the following relation

\begin{widetext}
\begin{equation}
    \begin{aligned}
        M^{\text{4-gon}}_{\ell}(1^{\pm}|2^{\pm}|3^{\pm}|4^{\pm})&= N^{\text{4-gon}}_{\ell}(1^{\pm}|2^{\pm}|3^{\pm}|4^{\pm})\times  N^{\text{4-gon}}_{\ell}(1^{\pm}|2^{\pm}|3^{\pm}|4^{\pm})\\
         M^{\text{3-gon}}_{\ell}(1^{\pm}2^{\pm}|3^{\pm}|4^{\pm})&=N^{\text{3-gon}}_{\ell}(1^{\pm}2^{\pm}|3^{\pm}|4^{\pm})\times  N^{\text{3-gon}}_{\ell}(1^{\pm}2^{\pm}|3^{\pm}|4^{\pm})\\
        M^{\text{2-gon}}_{\ell}(1^{\pm}2^{\pm}|3^{\pm}4^{\pm})&=N^{\text{2-gon}}_{\ell}(1^{\pm}2^{\pm}|3^{\pm}4^{\pm})\times  N^{\text{2-gon}}_{\ell}(1^{\pm}2^{\pm}|3^{\pm}4^{\pm})\\
    \end{aligned}
\end{equation}
\end{widetext}
One can detect this double copy structure from the contraction in (\ref{eq:107}) compared to (\ref{eq:57}), as the former one will recursively produce two copies of the last one. One can argue this formally by induction, beginning with the case $O_{n}=1234$, so we see that each of the contractions of the preintegrands $\mathcal{K}^{\pm\alpha\beta}_{\mu\nu T}$ will reproduce a contraction of a pair of currents $\mathcal{J}^{\pm\alpha}_{\mu T}\mathcal{J}^{\pm\beta}_{\nu T}$. Thus, after taking all possible deconcatenations, we will find a \emph{pair} of (anti-)self-dual Yang-Mills numerators:
\begin{widetext}
\begin{equation}
\begin{split}
    I^{\text{$1$-loop}}_{n}(\ell)&=\sum_{k=2}^{n}\sum_{O_{n}=P_{1}\cdots P_{k}}I^{\text{$k$-gon}}_{\ell}( P_{1}^{  \pm}|\cdots |P_{k}^{  \pm})\\
    &=\sum_{k=2}^{n}\sum_{O_{n}=P_{1}\cdots P_{k}}\frac{M^{\text{$k$-gon} }_{\ell}( P_{1}^{  \pm}|\cdots |P_{k}^{  \pm}) }{(\ell+ P_{1}+\cdots+ P_{k-1})^2\cdots (\ell + P_{1}  )^2 \ell^2 }\\
    &=\sum_{k=2}^{n}\sum_{O_{n}=P_{1}\cdots P_{k}}\frac{N^{\text{$k$-gon}}_{\ell}( P_{1}^{  \pm}|\cdots |P_{k}^{  \pm}) N^{\text{$k$-gon} }_{\ell}( P_{1}^{  \pm}|\cdots |P_{k}^{  \pm}) }{(\ell+ P_{1}+\cdots+ P_{k-1})^2\cdots (\ell + P_{1}  )^2 \ell^2}
    \end{split}
\end{equation}
\end{widetext}
This relation shows one-loop on-shell double copy for the self-dual gravity numerators obtained as a consequence of the off-shell perturbiner expansion for both self-dual Yang-Mills and gravity. 

\section{Conclusions}\label{section:concl}
We have provided a detailed description of the perturbative expansion for the (anti)self-dual sectors of Yang-Mills and gravity up to one-loop. Starting with the Yang-Mills sectors, the numerators for the currents obtained using the perturbiner method exhibit a fully off-shell version of color-kinematics duality. This property has been utilized to study various aspects, including the KLT kernel for currents and amplitudes, as well as to construct one-loop color-kinematics master numerators for any number of external legs. Future work will explore the calculation of master numerators for higher loop levels, building upon this special off-shell color-kinematics duality.

\begin{acknowledgments}
We thank Renann Lipinski Jusinskas and Joris Raeymaekers for fruitful discussions about this work. The work of CLA was partially supported by the European Structural and Investment Funds and the Czech Ministry of Education, Youth and Sports (project
FORTE CZ.02.01.01/00/22$\_$008/0004632).
\end{acknowledgments}

\nocite{*}


\medskip

\bibliographystyle{apsrev4-2}
\bibliography{REFERENCES}
\addcontentsline{toc}{section}{Bibliography}

\end{document}